\documentclass[twocolumn]{aastex631}
\usepackage{amsmath}

\newcommand{\jwst}{{\em JWST}}
\newcommand\titlelowercase[1]{\texorpdfstring{\lowercase{#1}}{#1}}

\makeatletter
\newcommand\footnoteref[1]{\protected@xdef\@thefnmark{\ref{#1}}\@footnotemark}
\makeatother

\begin{document}

\title{Evolution of the Mass--Metallicity Relation from Redshift $z\approx8$ to the Local Universe}

\correspondingauthor{Danial Langeroodi}
\email{danial.langeroodi@nbi.ku.dk}

\author[0000-0001-5710-8395]{Danial Langeroodi}
\affil{DARK, Niels Bohr Institute, University of Copenhagen, Jagtvej 128, 2200 Copenhagen, Denmark}

\author[0000-0002-4571-2306]{Jens Hjorth}
\affil{DARK, Niels Bohr Institute, University of Copenhagen, Jagtvej 128, 2200 Copenhagen, Denmark}
\author[0000-0003-1060-0723]{Wenlei Chen}
\affil{Minnesota Institute for Astrophysics, University of Minnesota, 116 Church Street SE, Minneapolis, MN 55455, USA}
\author[0000-0002-0786-7307]{Patrick L. Kelly}
\affil{Minnesota Institute for Astrophysics, University of Minnesota, 116 Church Street SE, Minneapolis, MN 55455, USA}
\author[0000-0002-1681-0767]{Hayley Williams}
\affil{Minnesota Institute for Astrophysics, University of Minnesota, 116 Church Street SE, Minneapolis, MN 55455, USA}
\author[0000-0001-8792-3091]{Yu-Heng Lin}
\affil{Minnesota Institute for Astrophysics, University of Minnesota, 116 Church Street SE, Minneapolis, MN 55455, USA}
\author[0000-0002-9136-8876]{Claudia Scarlata}
\affil{Minnesota Institute for Astrophysics, University of Minnesota, 116 Church Street SE, Minneapolis, MN 55455, USA}
\author[0000-0002-0350-4488]{Adi Zitrin}
\affil{Physics Department, Ben-Gurion University of the Negev, P.O. Box 653, Beer-Sheva 8410501, Israel}
\author[0000-0002-8785-8979]{Tom Broadhurst}
\affiliation{Department of Physics, University of the Basque Country UPV/EHU, E-48080 Bilbao, Spain}
\affiliation{DIPC, Basque Country UPV/EHU, E-48080 San Sebastian, Spain}
\affiliation{Ikerbasque, Basque Foundation for Science, E-48011 Bilbao, Spain}
\author[0000-0001-9065-3926]{Jose M. Diego}
\affil{IFCA, Instituto de F\'isica de Cantabria (UC-CSIC), Av. de Los Castros s/n, 39005 Santander, Spain}
\author[0000-0001-8156-0330]{Xiaosheng Huang}
\affil{Department of Physics \& Astronomy, University of San Francisco, San Francisco, CA 94117, USA}
\affil{Physics Division, Lawrence Berkeley National Laboratory, 1 Cyclotron Road, Berkeley, CA 94720, USA}
\author[0000-0003-3460-0103]{Alexei V. Filippenko}
\affil{Department of Astronomy, University of California, Berkeley, CA 94720-3411, USA}
\author[0000-0002-2445-5275]{Ryan J. Foley}
\affil{Department of Astronomy and Astrophysics, UCO/Lick Observatory, University of California, 1156 High Street, Santa Cruz, CA 95064, USA}
\author[0000-0001-8738-6011]{Saurabh Jha}
\affil{Department of Physics and Astronomy, Rutgers, The State University of New Jersey, Piscataway, NJ 08854, USA}
\author[0000-0002-6610-2048]{Anton M. Koekemoer}
\affil{Space Telescope Science Institute, 3700 San Martin Dr., Baltimore, MD 21218, USA}
\author[0000-0003-3484-399X]{Masamune Oguri}
\affil{Center for Frontier Science, Chiba University, 1-33 Yayoi-cho, Inage-ku, Chiba 263-8522, Japan}
\affil{Department of Physics, Chiba University, 1-33 Yayoi-Cho, Inage-Ku, Chiba 263-8522, Japan}
\author[0000-0002-2807-6459]{Ismael Perez-Fournon}
\affil{Instituto de Astrofisica de Canarias (IAC), E-38205 La Laguna, Tenerife, Spain}
\affil{Departamento de Astrof\'{\i}sica, Universidad de La Laguna (ULL), 38206 La Laguna, Tenerife, Spain}
\author[0000-0002-2361-7201]{Justin Pierel}
\affil{Space Telescope Science Institute, 3700 San Martin Dr., Baltimore, MD 21218, USA}
\author[0000-0002-5391-5568]{Frederick Poidevin}
\affil{Instituto de Astrofisica de Canarias (IAC), E-38205 La Laguna, Tenerife, Spain}
\affil{Departamento de Astrof\'{\i}sica, Universidad de La Laguna (ULL), 38206 La Laguna, Tenerife, Spain}
\author[0000-0002-7756-4440]{Lou Strolger}
\affil{Space Telescope Science Institute, 3700 San Martin Dr., Baltimore, MD 21218, USA}
\begin{abstract}

A tight positive correlation between the stellar mass and the gas-phase metallicity of galaxies has been observed at low redshifts. The redshift evolution of this correlation can strongly constrain theories of galaxy evolution. The advent of \jwst\ allows probing the mass-metallicity relation at redshifts far beyond what was previously accessible. Here we report the discovery of two emission-line galaxies at redshifts 8.15 and 8.16 in \jwst\ NIRCam imaging and NIRSpec spectroscopy of targets gravitationally lensed by the cluster RX\,J2129.4$+$0005. We measure their metallicities and stellar masses along with nine additional galaxies at $7.2 < z_{\textnormal{spec}} < 9.5$ to report the first quantitative statistical inference of the mass--metallicity relation at $z\approx8$. We measure $\sim 0.9$ dex evolution in the normalization of the mass--metallicity relation from $z \approx 8$ to the local Universe; at fixed stellar mass, galaxies are 8 times less metal enriched at $z \approx 8$ compared to the present day. Our inferred normalization is in agreement with the predictions of the FIRE simulations. Our inferred slope of the mass--metallicity relation is similar to or slightly shallower than that predicted by FIRE or observed at lower redshifts. We compare the $z \approx 8$ galaxies to extremely low metallicity analog candidates in the local Universe, finding that they are generally distinct from extreme emission-line galaxies or ``green peas" but are similar in strong emission-line ratios and metallicities to ``blueberry galaxies". Despite this similarity, at fixed stellar mass, the $z \approx 8$ galaxies have systematically lower metallicities compared to blueberry galaxies.

\end{abstract}



\section{Introduction} \label{sec:intro}

The gas-phase metallicity of a galaxy measures its current state of chemical enrichment, holding a record of its star-formation history (SFH), gas infall, feedback, and merger history. 
These mechanisms are not identical for galaxies of different stellar mass at a given redshift, as evidenced by the positive empirical correlation between the gas-phase metallicity and stellar mass: the mass--metallicity relation \citep{1968JRASC..62..145V, 1970A&A.....7..311P, 1979A&A....80..155L}.
This correlation has been extensively studied in the local Universe with numerous works deriving a tight mass--metallicity relation that spans five decades of stellar mass from $10^7\, M_{\odot}$ to $10^{12}\, M_{\odot}$ and only starts to saturate in metallicity at $M_{\star} > 10^{10}\, M_{\odot}$ \citep{2004ApJ...613..898T, 2006ApJ...647..970L, 2006ApJ...636..214V, 2008ApJ...681.1183K, 2010MNRAS.408.2115M, 2012ApJ...754...98B, 2013A&A...549A..25P, 2013MNRAS.432.1217P, 2013ApJ...765..140A, 2013ApJ...765...66H, 2015ApJ...800..121H, 2015MNRAS.446.1449L, 2016ApJ...828...67L, 2019ApJ...877....6B, maiolino+2019, curti+2020b, sanders+2021}. 

Beyond the local Universe the mass--metallicity relation has been inferred out to $z \approx 3.5$, showing the same general trends as seen in the local Universe but with a lower normalization; galaxies of the same stellar mass at higher redshifts seem to be less chemically enriched \citep{2005ApJ...635..260S, 2006ApJ...644..813E, 2008A&A...488..463M, 2009MNRAS.398.1915M, 2011ApJ...730..137Z, 2014ApJ...791..130Z, 2014ApJ...792...75Z, 2012ApJ...755...73W, 2016ApJ...827...74W, 2013ApJ...772..141B, 2013ApJ...776L..27H, 2013ApJ...774..130K, 2014MNRAS.440.2300C, 2014MNRAS.437.3647Y, 2014ApJ...792....3M, 2014ApJ...795..165S, 2014A&A...563A..58T, 2016ApJ...826L..11K, 2015ApJ...802L..26K, 2015ApJ...805...45L, 2016ApJ...828...67L, 2015ApJ...799..138S, 2018ApJ...858...99S, 2020ApJ...888L..11S, sanders+2021, 2016MNRAS.463.2002H, 2016ApJ...822...42O, 2017ApJ...849...39S}.
The inference of a mass--metallicity relation beyond $z = 3.5$ has been stalled thus far because the primary rest-optical metallicity indicators get redshifted beyond near-infrared wavelengths where the bright sky background and the reduced atmospheric transmission prohibits emission-line measurements \citep[see][and references therein]{maiolino+2019, sanders+2021}. 

Despite these challenges, several groups have attempted to measure gas-phase metallicities at redshifts beyond 3.5 through alternative methods. 
\cite{2016ApJ...822...29F} used a calibration of the rest-ultraviolet (UV) absorption lines to estimate the metallicity in three mass bins at $z \approx 5$ from stacked spectra of a sample of $3.5 < z < 6.0$ galaxies, detected in the cosmic evolution survey \citep[COSMOS;][]{COSMOS} and spectroscopically confirmed with the Deep Imaging Multi-object Spectrograph \citep[DEIMOS;][]{2003SPIE.4841.1657F}. 
\cite{jones+2020} presented a calibration of the ALMA-accessible (the Atacama Large Millimeter/submillimeter Array) far-infrared [\ion{O}{3}]$\lambda88\mu$m emission-line intensity as a direct-method (i.e., calibrated against the ``direct $T_e$ method") metallicity estimator, and used it to measure the metallicities of a sample of six galaxies (five with mass measurements) at $z \approx 8$. 
However, these studies could not significantly constrain the mass--metallicity relation at high redshifts because of the small sample size and large statistical and/or systematic uncertainties \citep[see][for a discussion on systematic uncertainties]{maiolino+2019}.

Tuned to reproduce the mass--metallicity relation at $z < 3.5$, theoretical models and simulations of galaxy evolution have predicted the shape and normalization of the mass--metallicity relation at higher redshifts. \cite{ma+2016} inferred the mass--metallicity relation and its evolution up to $z = 6 $ from the FIRE simulations and demonstrated reasonable agreement with the observed relation and its evolution up to $z = 3$ for a broad range in stellar mass. \cite{ma+2016} concluded that the redshift evolution of the mass--metallicity relation coincides with the redshift evolution of the stellar mass fraction \citep[see, e.g.,][]{UM1, UM2, UM3}, potentially pointing toward a universal relation between the stellar mass, gas mass, and metallicities. Although pending empirical confirmation, their results can be extrapolated to redshifts beyond the current observational limits. Similar conclusions have been made based on EAGLE \citep{2015MNRAS.446..521S, 2016MNRAS.459.2632L, 2017MNRAS.472.3354D}, Illustris TNG \citep{2019MNRAS.484.5587T}, and FirstLight \citep{2020MNRAS.494.1988L} simulations and reproduced by semi-analytic models \citep[see, e.g.,][]{2016MNRAS.461.1760H, 2023MNRAS.518.3557U}.


Furthermore, theoretical models and simulations have identified the stellar and active galactic nucleus (AGN) feedback-driven outflows, the metal content of the outflows in comparison to the interstellar medium (ISM), the shape and evolution of the stellar initial mass function (IMF), and the dependency of stellar yields on redshift and galaxy stellar mass as the primary drivers of shape and normalization of the mass--metallicity relation \citep[see, e.g.,][and references therein]{2018MNRAS.474.1143L}. 
Probing the mass--metallicity relation at $z \approx 8$ and beyond is of critical importance in characterizing the mechanisms deriving the shape and redshift evolution of the mass--metallicity relation, because this is the epoch when galaxies are expected to have much simpler star-formation histories (SFHs), feedback histories, and merger histories which allow for a more robust comparison with galaxy evolution theoretical models and simulations.

The NIRSpec instrument \citep{2022A&A...661A..80J} onboard \jwst\ has already demonstrated tremendous capability in spectroscopically confirming the high-redshift NIRCam-selected candidates with relative ease \citep[see, e.g.,][]{carnall+2022, williams+2022, 2022arXiv221015639R, 2022arXiv221109097M}.
For the first time, NIRSpec enables high signal-to-noise ratio (S/N) detections of the rest-frame optical metallicity diagnostic emission lines with high spectral resolution; this has resulted in ``direct $T_e$" or ``strong line" metallicity measurements of a growing sample of galaxies at $z \approx 8$ and beyond \citep[see, e.g.,][]{curti+2022, schaerer+2022, williams+2022}. 

In this work, we present the discovery of two galaxies detected in the field of the foreground lensing cluster RX\,J2129.4$+$0005, in imaging and spectroscopy acquired as part of a Director's Discretionary program (DD-2767; PI P. Kelly) to observe a strongly lensed background supernova. They have spectroscopic redshifts of $z =  8.16$ (RX2129--ID11002) and 8.15 (RX2129--ID11022), based on emission lines detected with NIRSpec prism observations. We obtain gas-phase metallicity measurements for these galaxies using rest-frame optical emission-line metallicity indicators. 
We combine these new measurements with literature \jwst\ and ALMA metallicity measurements of galaxies at $z \approx 8$ to construct a sample of eleven galaxies with ``direct $T_e$", ``strong line", or far-infrared emission line metallicity measurements at this redshift. We measure the stellar masses of the entire sample of eleven galaxies, and for the first time significantly constrain both the slope and the normalization of the mass--metallicity relation at $z \approx 8$ as well as the evolution of its normalization from $z \approx 8$ to the present day.

Young, low-metallicity galaxies in the nearby universe have been proposed as analogs of high-redshift galaxies. In particular, the so-called ``extreme emission-line galaxies" (EELGs), ``green peas," and ``blueberry galaxies" are interesting candidates for having properties similar to those that are being revealed at high redshift. EELGs \citep{eelgs} were identified in zCOSMOS as $z < 1$ galaxies with strong emission lines; higher redshift EELGs ($z\approx3$) were proposed to be analogs of very high redshift galaxies \citep{2017NatAs...1E..52A}. Green peas \citep{2009MNRAS.399.1191C, yang+2017g} are compact SDSS galaxies with strong [\ion{O}{3}] in the range $0.14<z<0.36$; their properties are very similar to those of EELGs. Blueberry galaxies are similar to green peas, but are selected to be at low redshifts ($z<0.05$) and hence probe fainter luminousities and lower stellar masses \citep{yang+2017b}. The first three \jwst\ NIRSpec-identified galaxies at $z\approx 8$ have been discussed in this context and have been likened individually to green peas or blueberry galaxies \citep{schaerer+2022,2022ApJ...939L...3T,2022arXiv220713020R,2023MNRAS.518..592K}.
For the six \jwst\ detected galaxies at $z\approx8$, we find that their emission-line properties are very similar to those of blueberry galaxies as a population. However, we find that the $z \approx 8$ galaxies stand out from the blueberry galaxies in the mass--metallicity diagram. At a given metallicity, $z\approx8$ galaxies have higher stellar masses than blueberry galaxies or green peas. 

Throughout this work we adopt a standard $\Lambda$CDM cosmology with H$_{0} = 70$ km s$^{-1}$ Mpc$^{-1}$, $\Omega_\textnormal{m}$ = 0.3, and $\Omega_{\Lambda}$ = 0.7. Furthermore, we adopt a \cite{chabrier+2003} stellar IMF, and magnitudes are in the AB system \citep{1983ApJ...266..713O}. 


\begin{figure*}
    \centering
    \includegraphics[width=18cm]{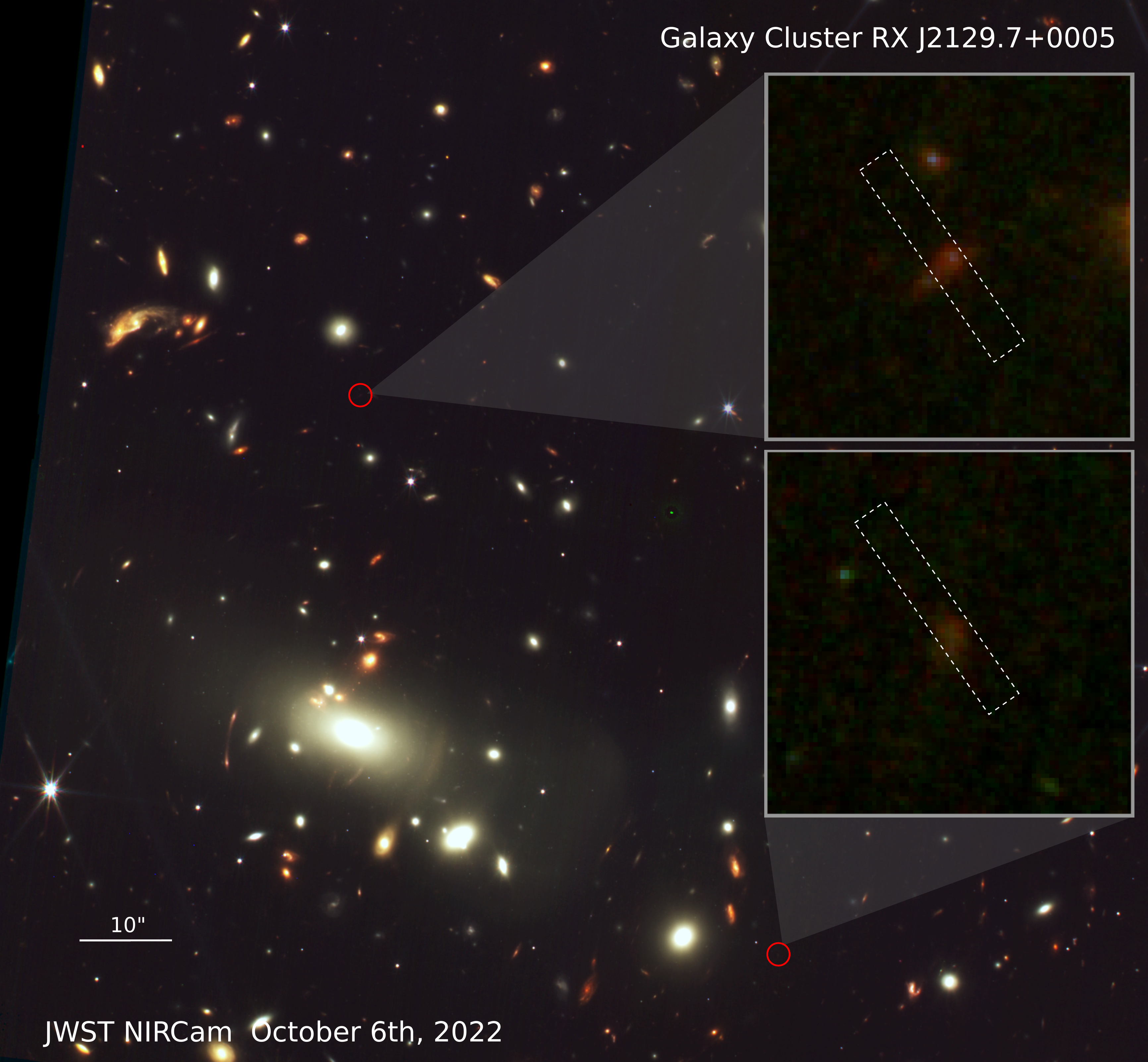}
    \caption{NIRCam color-composite image of the RX\,J2129.4$+$0005 lensing cluster (R: F356W+F444W, G: F200W+F277W, B: F115W+F150W). The two $z \approx 8.15$ galaxies are indicated by red circles; insets show the positions of the NIRSpec MSA slits. The upper inset shows the photometry of RX2129--ID11002 at $z_{\textnormal{spec}} = 8.16$ (RA(J2000.0) = 21:29:39.904, Dec(J2000.0) = 00:05:58.83) and the lower inset shows RX2129--ID11022 at $z_{\textnormal{spec}} = 8.15$ (RA(J2000.0) = 21:29:36.080, Dec(J2000.0) = 00:04:56.53). We do not expect these galaxies to be multiply imaged; the lensing magnifications are presented in Table \ref{table: line fluxes}.}
    \label{fig: photometry}
\end{figure*}

\section{Sample \titlelowercase{of $z \approx 8$} galaxies} \label{sec: sample}

With the addition of strong line metallicity\footnote{Unless otherwise specified, throughout this work metallicity refers to the gas-phase metallicity of galaxies.} measurements for the RX2129--ID11002 and RX2129--ID11022 galaxies presented in this work (see Section \ref{sec: metallicity} for metallicity measurements), we can construct a sample of eleven $z \approx 8$ galaxies with available multi-band photometry and metallicities measured through either the direct $T_e$ method or other empirical methods calibrated against this method (i.e., the strong line method and the \cite{jones+2020} calibration of the [\ion{O}{3}]$\lambda88\mu$m emission-line intensity; see Section \ref{sec: metallicity} for more details on both methods). This sample includes the RX2129--ID11027 galaxy presented in \cite{williams+2022}, the three galaxies detected in the field towards the SMACS\,J0723.3$-$7327 galaxy cluster \citep{carnall+2022}, and the pre-\jwst\ sample compiled by \cite{jones+2020}. In this Section we provide an overview of this sample. The spectroscopic redshifts, magnification factors, and original literature references for the entire sample are available in Table \ref{table: mass}.

\subsection{RX2129 galaxies}

\begin{figure*}
    \centering
    \includegraphics[width=18cm]{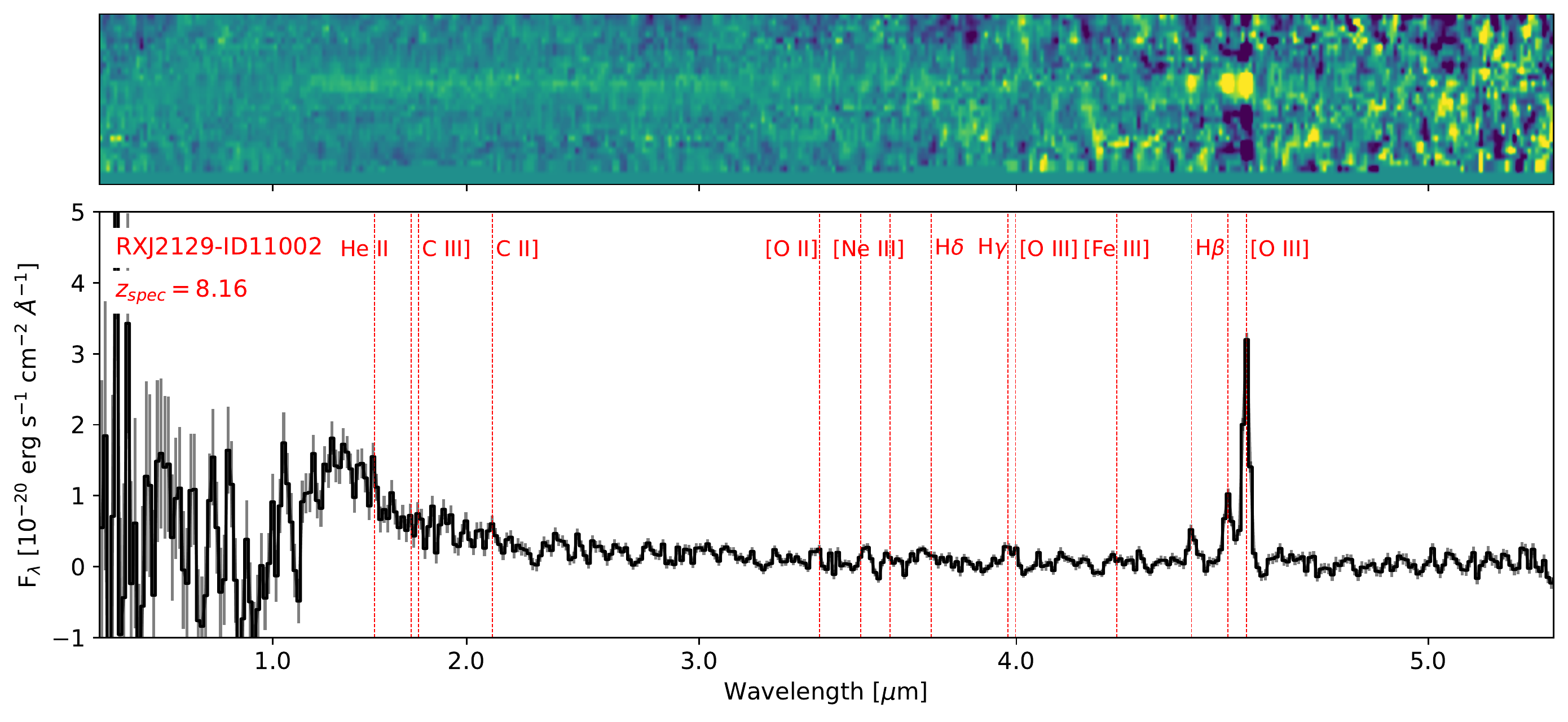}
    \caption{NIRSpec 2D (top panel) and 1D (bottom panel) spectra of the RX2129--ID11002 galaxy at $z_{\textnormal{spec}} = 8.16$. The red dashed lines show the identified emission lines, and the thin grey lines show the $1\sigma$ uncertainties. The fits to the emission lines and continuum are shown in Figure \ref{fig: decomposition 11002}, and the emission line flux measurements are presented in Table \ref{table: line fluxes}. The spectrum is not corrected for lensing magnification.}
    \label{fig: spectra 11002}
\end{figure*}

\begin{figure*}
    \centering
    \includegraphics[width=18cm]{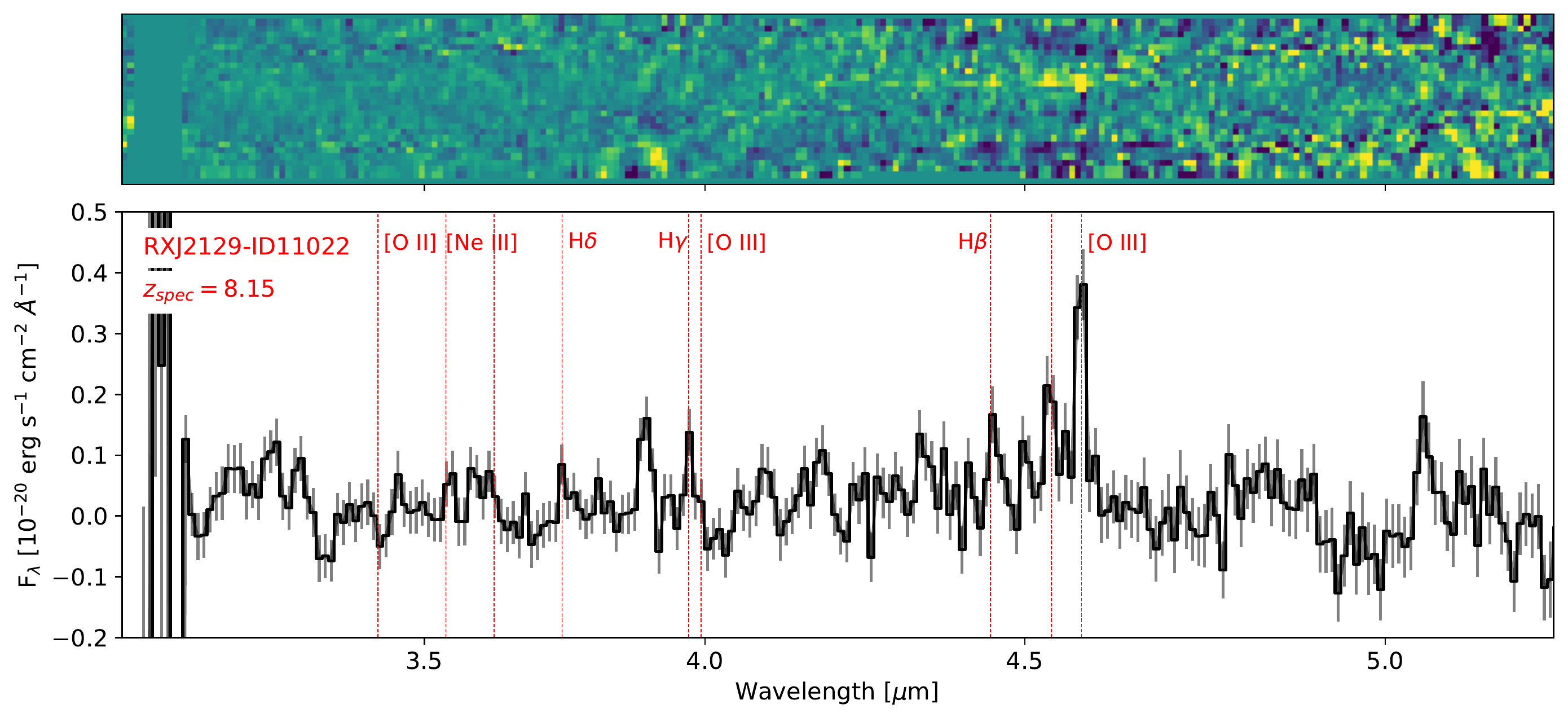}
    \caption{NIRSpec 2D (top panel) and 1D spectra (bottom panel) of the RX2129--ID11022 galaxy at $z_{\textnormal{spec}} = 8.15$. The red dashed lines show the identified emission lines, and the thin grey lines show the $1\sigma$ uncertainties. The fits to the emission lines and continuum are shown in Figure \ref{fig: decomposition 11022}, and the emission line flux measurements are presented in Table \ref{table: line fluxes}. The spectrum is not corrected for lensing magnification. The blue part of the spectrum is missing for this galaxy because the NIRSpec MSA configuration caused the spectrum to only partially fall on the detector.}
    \label{fig: spectra 11022}
\end{figure*}

\begin{figure*}
    \centering
    \includegraphics[width=16.5cm]{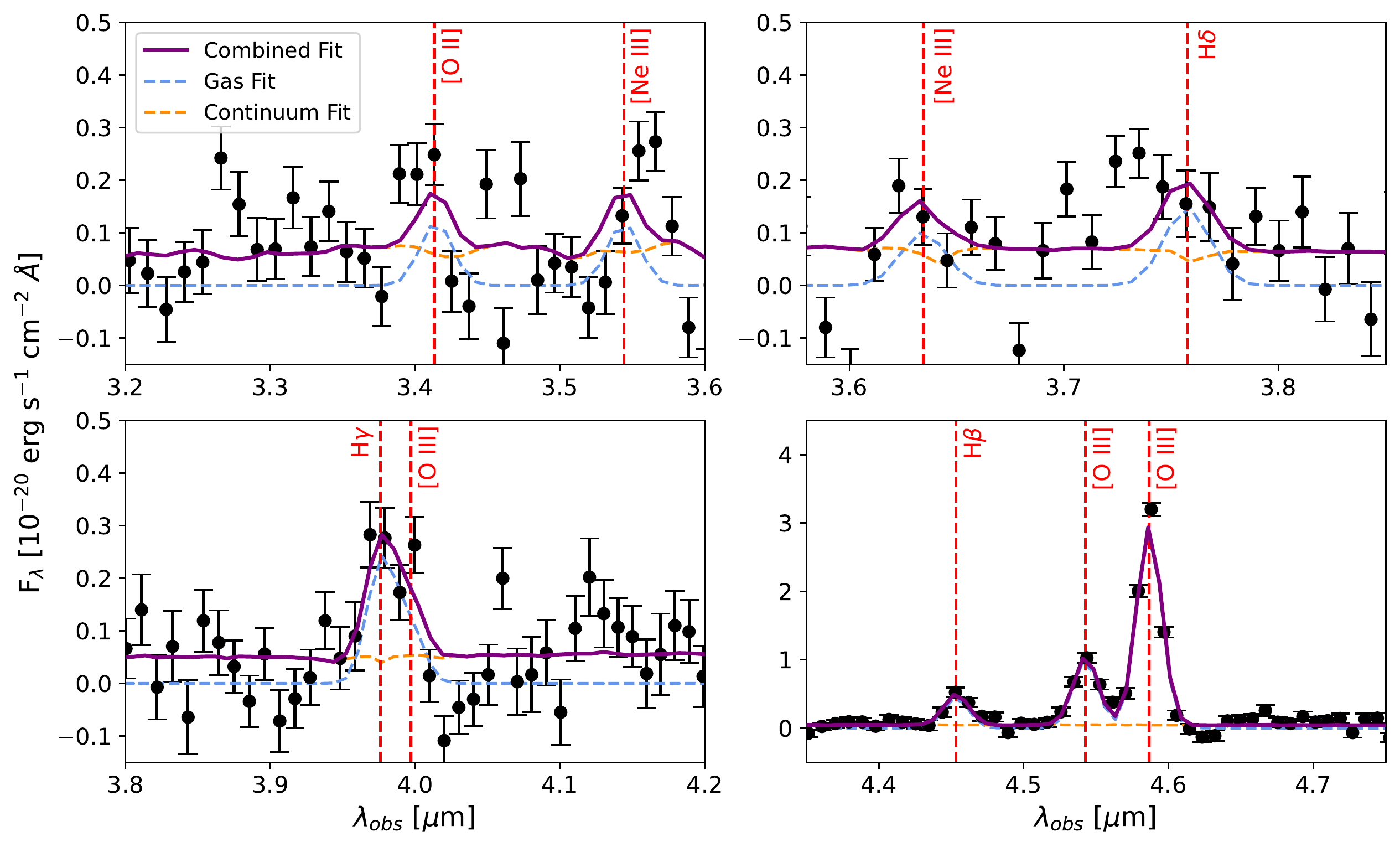}
    \caption{Emission lines fits (using \texttt{pPXF}) to the NIRSpec 1D spectrum of the RX2129--ID11002 (see Figure \ref{fig: spectra 11002}) galaxy at $z_{\textnormal{spec}} = 8.16$. The dashed orange line shows the stellar continuum fit and the dashed blue line shows the Gaussian fits to the emission lines; the solid purple line shows the combined continuum plus emission lines fit. The line widths are fixed to the spectral resolution of NIRSpec prism at the observed wavelength. The measured emission line fluxes are reported in Table \ref{table: line fluxes}.}
    \label{fig: decomposition 11002}
\end{figure*}

\begin{figure*}
    \centering
    \includegraphics[width=16.5cm]{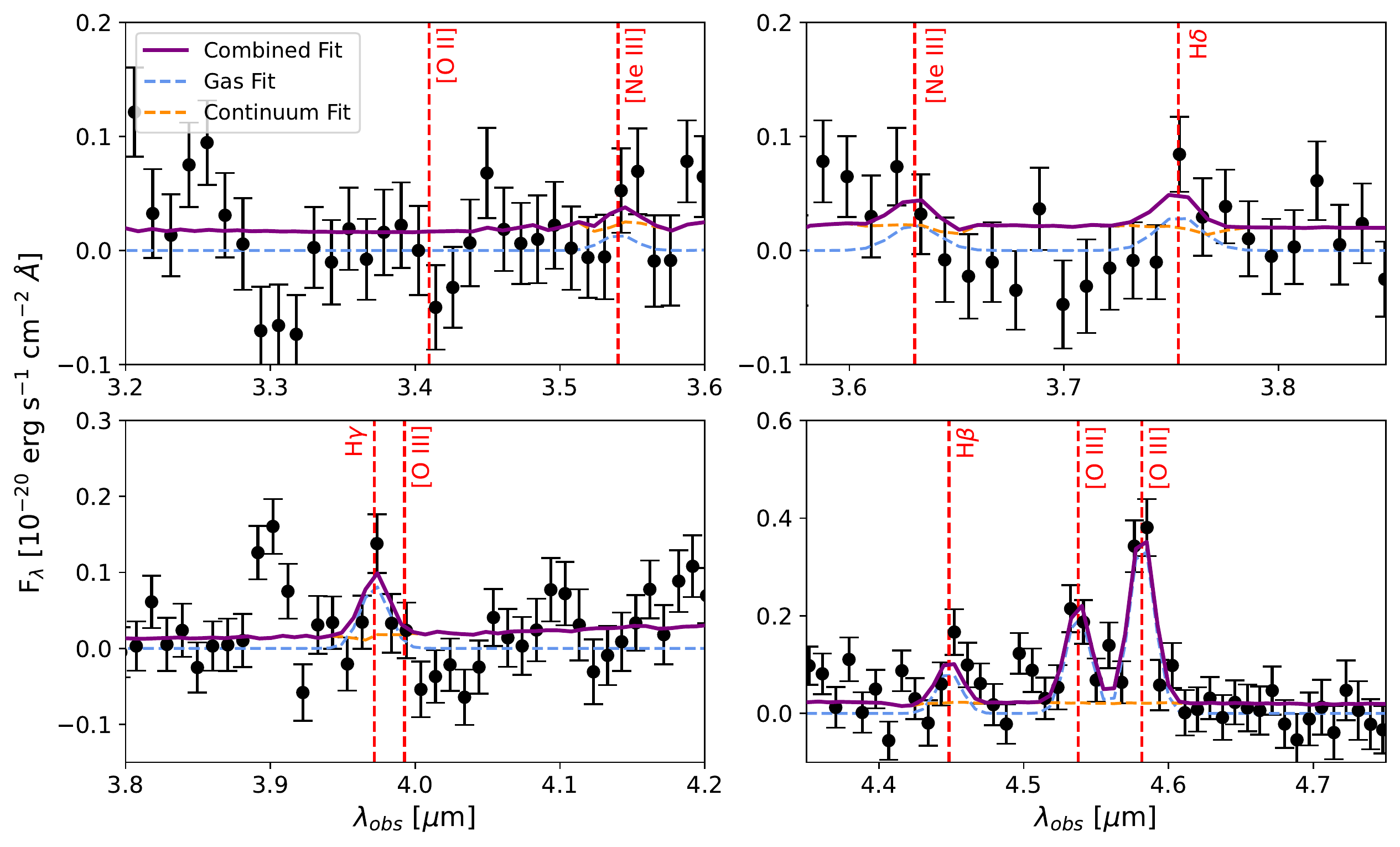}
    \caption{Same as Figure \ref{fig: decomposition 11002}, but for the RX2129--ID11022 galaxy at $z_{\textnormal{spec}} = 8.15$. The 1D spectrum is shown in Figure \ref{fig: spectra 11022}.}
    \label{fig: decomposition 11022}
\end{figure*}

The imaging of the RX\,J2129.4$+$0005 galaxy cluster (RX2129 for short) was obtained as part of the DD-2767 program (PI: Kelly) with the \jwst\ NIRCam instrument in the F115W, F150W, F200W, F277W, F356W, and F444W filters. We present a color-composite image in Figure \ref{fig: photometry}. The details of NIRCam observations and data reduction are presented in \cite{williams+2022}. Spectra for a sample of high redshift galaxy candidates, identified using the \texttt{EAZY} \citep{brammervandokkumcoppi08} photometric redshift estimation code, were subsequently obtained with the \jwst\ NIRSpec instrument as part of the same DD program. The NIRSpec spectra were obtained in multi-object spectroscopy mode with the prism disperser, which provides wavelength coverage from 0.6 $\mu$m to 5.3 $\mu$m. The spectral resolution ranges from $R\approx 50$ at the blue end to $R \approx 400$ at the red end. Based on this spectra, three candidates were confirmed at $z_{\textnormal{\scriptsize spec}} > 8$: RX2129--ID11002, RX2129--ID11022, and RX2129--ID11027.

The photometry of RX2129--ID11002 and RX2129--ID11022 are presented in Table \ref{table: photometry}; their color-composite images are presented in the smaller panels of Figure \ref{fig: photometry}. We measure the lensing magnification of these galaxies based on the model presented in \cite{williams+2022} \citep[see also][]{caminha+2019, 2021MNRAS.508.1206J}. 
This model is constructed using \texttt{glafic} \citep{oguri+2010, oguri+2021}, and the measured magnifications are further confirmed with the Zitrin-parametric code \citep{zitrin+2015}. The magnification factors are reported in Table \ref{table: line fluxes}. Based on these models we do not expect RX2129--ID11002 or RX2129--ID11022 to be multiply imaged.

We also present the NIRSpec spectra of RX2129--ID11002 and RX2129--ID11022 and establish their spectroscopic redshifts. The NIRSpec data of RX2129--ID11002 and RX2129--ID11022 were reduced following the method described in \cite{williams+2022}; we measure $z_{\textnormal{\scriptsize spec}} = 8.16 \pm 0.01$ and $z_{\textnormal{\scriptsize spec}} = 8.15 \pm 0.01$, respectively. The reduced spectra of these galaxies are presented in Figures \ref{fig: spectra 11002} and \ref{fig: spectra 11022}. We present their strong line analysis and metallicity measurements in Section \ref{sec: strong line}.

The NIRCam photometry and NIRSpec spectra of the RX2129--ID11027 galaxy were reduced and presented in \cite{williams+2022}, measuring $z_{\textnormal{\scriptsize spec}} = 9.51 \pm 0.01$. \cite{williams+2022} also reported the strong line analysis and metallicity measurement of this galaxy. The ionization properties of all three RX2129 galaxies, including their UV magnitudes, UV slopes, escape fractions of ionizing radiation, and ionizing photon production efficiencies are reported in \cite{2023arXiv230304572L}.

\begin{deluxetable}{llll}
\tablewidth{0pt}
\tablecaption{Measured NIRCam photometry (and $1\sigma$ uncertainty) of the RX2129--ID11002 and RX2129--ID11022 galaxies, in AB magnitudes.}
\label{table: photometry}
\tablehead{
\colhead{Filter} &
\colhead{$\lambda$(\AA)} &
\colhead{RX2129--ID11002} &
\colhead{RX2129--ID11022} 
}
\startdata
F115W & 11543.01 & 27.540 $\pm$ 0.299 & 31.627 $\pm$ 6.190\\ 
F150W & 15007.45 & 26.849 $\pm$ 0.090 & 28.481 $\pm$ 0.175\\ 
F200W & 19886.48 & 27.397 $\pm$ 0.180 & 29.614 $\pm$ 0.693\\ 
F277W & 27577.96 & 27.033 $\pm$ 0.104 & 28.539 $\pm$ 0.229\\ 
F356W & 35682.28 & 26.885 $\pm$ 0.072 & 29.472 $\pm$ 0.304\\ 
F444W & 44036.71 & 26.124 $\pm$ 0.077 & 28.120 $\pm$ 0.212\\ 
\enddata
\end{deluxetable}

\subsection{SMACS0723 galaxies}

The NIRCam and MIRI imaging as well as the NIRSpec multi-object spectroscopy of the SMACS\,J0723.3$-$7327 galaxy cluster (SMACS0723 for short) were obtained as part of the \jwst\ Early Release Observations \citep{ERO}. The cluster was observed in the NIRCam F090W, F150W, F200W, F277W, F356W, and F444W filters and MIRI F770W, F1000W, F1500W, and F1800W filters. \cite{carnall+2022} analysed the spectra and measured secure redshifts for 10 galaxies (out of the total available 35 objects), 3 of which turned out to be at $z \approx 8$: SMACS0723--ID4590 ($z_{\textnormal{\scriptsize spec}} = 8.498$), SMACS0723--ID6355 ($z_{\textnormal{\scriptsize spec}} = 7.665$), and SMACS0723--ID10612 ($z_{\textnormal{\scriptsize spec}} = 7.663$). As detailed in Section \ref{sec: metallicity}, for these galaxies we adopt the gas-phase metallicities reported in the literature.

\subsection{Pre-JWST sample} 

Among the $z \approx 8$ galaxies that have been spectroscopically confirmed prior to the launch of \jwst, metallicities for six galaxies have been measured by \cite{jones+2020} using the ALMA-measured intensity of the $[$O{\footnotesize\;III}$]\lambda88\mu$m emission line (see Section \ref{sec: metallicity} for more details). Multi-band photometry for five of these galaxies are available in the literature (see Table \ref{table: mass} for references). In the case of the BDF--3299 galaxy, the 6th galaxy in \cite{jones+2020} sample, there is a significant spatial offset between the far-infrared emission line and the rest-frame UV continuum. Therefore the measured metallicity does not correspond to the same region probed by photometry. Moreover, this galaxy has been detected in only one photometry band \citep{2011ApJ...730L..35V} which is not sufficient for an accurate stellar mass measurement; therefore we do not include it in our sample.

\section{Strong line analysis} \label{sec: strong line}

\subsection{Line intensity measurement}

\begin{deluxetable}{lll}
\tablewidth{0pt}
\tablecaption{Strong emission line flux measurements from the NIRSpec 1D spectrum of the RX2129--ID11002 and RX2129--ID11022 galaxies (see Figures \ref{fig: spectra 11002} and \ref{fig: spectra 11022}). The flux and $1\sigma$ uncertainties are reported in units of $10^{-19}$ erg s$^{-1}$ cm$^{-2}$. These measurements are not corrected for lensing magnification. In the bottom row we report the gas-phase metallicities measured using the strong line method from \cite{izotov+2019} (see Section \ref{sec: metallicity}).}
\label{table: line fluxes}
\tablehead{
\colhead{Emission line $[$\AA$]$} &
\colhead{RX2129--ID11002} &
\colhead{RX2129--ID11022}
}
\startdata
$\textnormal{$[$O{\footnotesize\;II}$]\lambda\lambda 3727,3729$}$ & $3.51 \pm 1.82$ & $0.0 \pm 1.21$ \\ 
$\textnormal{$[$Ne{\footnotesize\;III}$]\lambda 3869$}$ & $3.96 \pm 1.64$ & $0.93 \pm 0.92$ \\ 
$\textnormal{$[$Ne{\footnotesize\;III}$]\lambda 3968$}$ & $2.09 \pm 1.13$ & $0.78 \pm 0.76$ \\ 
$\textnormal{H$\delta$}$ & $3.99 \pm 1.48$ & $1.07 \pm 0.81$ \\ 
$\textnormal{H$\gamma$}$ & $5.28 \pm 1.24$ & $1.96 \pm 0.95$ \\ 
$\textnormal{$[$O{\footnotesize\;III}$]\lambda 4363$}$ & $2.35 \pm 1.14$ & $0.0 \pm 0.85$ \\ 
$\textnormal{H$\beta$}$ & $9.42 \pm 1.15$ & $1.27 \pm 0.89$ \\ 
$\textnormal{$[$O{\footnotesize\;III}$]\lambda 4959$}$ & $20.51 \pm 1.16$ & $4.01 \pm 0.93$ \\ 
$\textnormal{$[$O{\footnotesize\;III}$]\lambda 5007$}$ & $62.92 \pm 1.54$ & $7.16 \pm 1.01$ \\ 
\hline
$\textnormal{Redshift}$ & 8.16 & 8.15 \\ 
\textnormal{Magnification\tablenotemark{a}} & 2.23 $\pm$ 0.15 & 3.29 $\pm$ 0.33 \\ 
$\textnormal{$12 + \log(\textnormal{O}/\textnormal{H})$}$ & $7.65 \pm 0.09\tablenotemark{b}$ & $7.72$\tablenotemark{c} \\ 
\enddata
\tablenotetext{a}{Based on the model constructed using \texttt{glafic} \citep{oguri+2010, oguri+2021}, but also confirmed using the Zitrin-parametric code \citep{zitrin+2015}; see \cite{williams+2022} for a detail description of our lens model.}
\tablenotetext{b}{Includes both the statistical and systematic $1\sigma$ uncertainty}
\tablenotetext{c}{$1\sigma$ upper limit}
\end{deluxetable}

We measure the intensity of emission lines in the NIRSpec 1D spectrum of RX2129--ID11002 and RX2129--ID11022 using the Penalized PiXel-Fitting package \citep[\texttt{pPXF;}][]{2004PASP..116..138C, 2017MNRAS.466..798C, 2022arXiv220814974C}. \texttt{pPXF} adopts a maximum penalized likelihood method \citep{1997AJ....114..228M} to subtract the stellar continuum by modelling it with a stellar population and measures the line fluxes by fitting them with Gaussian profiles. We use the same \texttt{pPXF} setup as described in \cite{williams+2022}, with the MILES stellar library \citep{MILES1, MILES2}. The \texttt{pPXF} Gaussian fits to the emission lines of RX2129--ID11002 and RX2129--ID11022 are shown in Figures \ref{fig: decomposition 11002} and \ref{fig: decomposition 11022}, respectively. The measured emission line fluxes are reported in Table \ref{table: line fluxes}.

The $[$O{\footnotesize\;III}$]\lambda \lambda 4959, 5007$\AA\ doublet is resolved in our NIRSpec spectra. We fit each of the $[$O{\footnotesize\;III}$]\lambda \lambda 4959$\AA\ and $[$O{\footnotesize\;III}$]\lambda 5007$\AA\ emission lines independently, measuring $[$O{\footnotesize\;III}$]\lambda \lambda 5007$\AA/$[$O{\footnotesize\;III}$]\lambda \lambda 4959$\AA\ flux ratios of $3.07 \pm 0.19$ and $1.79 \pm 0.48$ for RX2129--ID11002 and RX2129--ID11022, respectively. The measured ratio for RX2129--ID11002 is in very good agreement with the 2.98 value set by atomic physics \citep{2000MNRAS.312..813S}. While this is not the case for RX2129--ID11022, it can be explained by the much lower S/N of the observed spectrum of this galaxy. We investigate this further by re-running \texttt{pPXF} with the ratio of $[$O{\footnotesize\;III}$]\lambda \lambda 4959, 5007$\AA\ doublet fixed to the 2.98 value set by atomic physics. The measured flux of $[$O{\footnotesize\;III}$]\lambda \lambda 4959, 5007$\AA\ doublet as well as the strong line metallicity measurement for RX2129--ID11002 remain intact. This setup results in measuring $2 \pm 17\%$ less $[$O{\footnotesize\;III}$]\lambda \lambda 4959, 5007$\AA\ doublet flux for RX2129--ID11022, but only decreases the $1\sigma$ metallicity upper limit (see Section \ref{sec: metallicity}) for this galaxy by 0.21 dex. By re-fitting the mass-metallicity relation, we confirm that this does not affect its best-fit normalization and slope (see Section \ref{sec: MZ}).

In this Section we compare the emission line properties of our sample of $z \approx 8$ galaxies with those of extremely low-metallicity analog candidates in the local Universe. \cite{izotov+2019} suggested the use of two emission line diagnostic diagrams to select extremely low-metallicity galaxies\footnote{O32 is defined as $[$O{\footnotesize\;III}$]\lambda 5007$\AA/$[$O{\footnotesize\;II}$]\lambda \lambda 3727, 3729$\AA, and R23 as ($[$O{\footnotesize\;II}$]\lambda \lambda 3727,3729$\AA\ + $[$O{\footnotesize\;III}$]\lambda 4959$\AA\ + $[$O{\footnotesize\;III}$]\lambda 5007$\AA)/H$\beta$.}: $[$O{\footnotesize\;III}$]\lambda 5007$\AA/H$\beta$ vs $[$O{\footnotesize\;II}$]\lambda \lambda 3727, 3729$\AA/H$\beta$ and O32 vs (R23$-$0.08O32). This was motivated by their calibration of the ``strong line" metallicity measurement method, where metallicity is calculated as a function of O32 and R23 (see Equation \ref{eq: metallicity} and Section \ref{sec: metallicity}). Here, we compare the locations of $z \approx 8$ NIRSpec emission line galaxies on these diagnostic diagrams with those of the proposed low-metallicity local Universe analogs: EELGs, green peas, and blueberry galaxies. 

\begin{figure*}
    \centering
    \includegraphics[width=18cm]{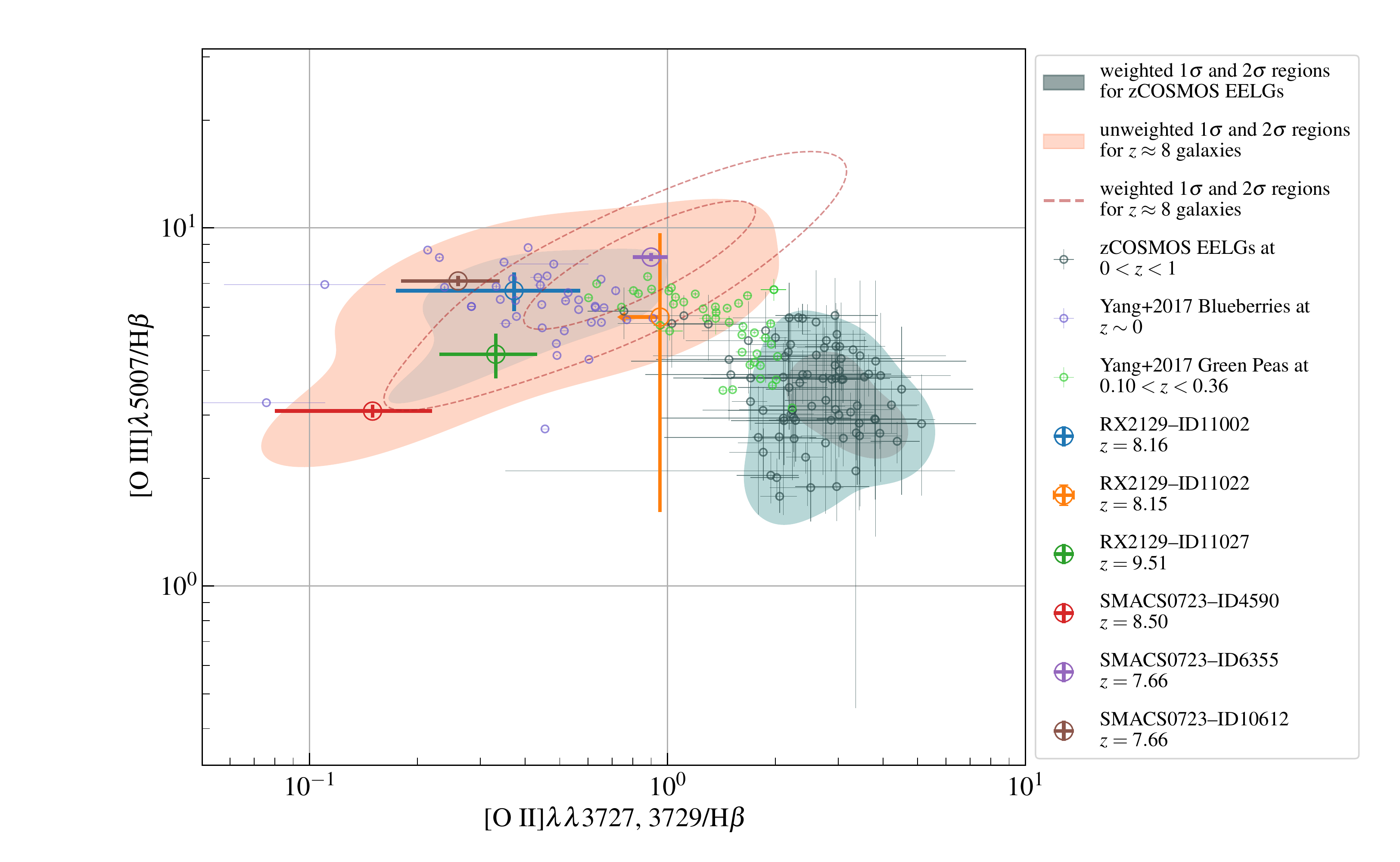}
    \caption{$[$O{\footnotesize\;III}$]\lambda 5007$\AA/H$\beta$ line flux ratio plotted against $[$O{\footnotesize\;II}$]\lambda \lambda 3727, 3729$\AA/H$\beta$ line flux ratio for 6 galaxies at $z_{\textnormal{\scriptsize spec}} \approx 8$ (large colored data points), as inferred from the \jwst\ NIRSpec observations of the RX2129 and SMACS0723 lensing clusters. The small dark-green data points show the measurements for the extreme emission line galaxies (EELGs) at $z_{\textnormal{\scriptsize spec}} \lesssim 1$ from the zCOSMOS survey. The dashed red contours indicate the weighted $1\sigma$ and $2\sigma$ confidence intervals for the $z \approx 8$ sample; since the weighted confidence intervals for this sample are dominated by the tight constraints on the SMACS0723--ID6355 galaxy we also show the unweighted $1\sigma$ and $2\sigma$ confidence intervals as the shaded red region. The shaded green region indicates the $1\sigma$ and $2\sigma$ confidence interval for the $z_{\textnormal{\scriptsize spec}} \lesssim 1$ EELGs sample. Moreover, we show the blueberry (small purple data points) and green pea (small light-green data points) galaxies from \cite{yang+2017b, yang+2017g}, confirming their remarkable strong emission line similarities (see also Figure \ref{fig: R23}) to the emission line-detected galaxies at $z \approx 8$; this is especially the case for the blueberry galaxies, almost all of which lie within the $2\sigma$ credible interval of the $z \approx 8$ galaxies.}
    \label{fig: OIII_to_OII}
\end{figure*}

\begin{figure*}
    \centering
    \includegraphics[width=18cm]{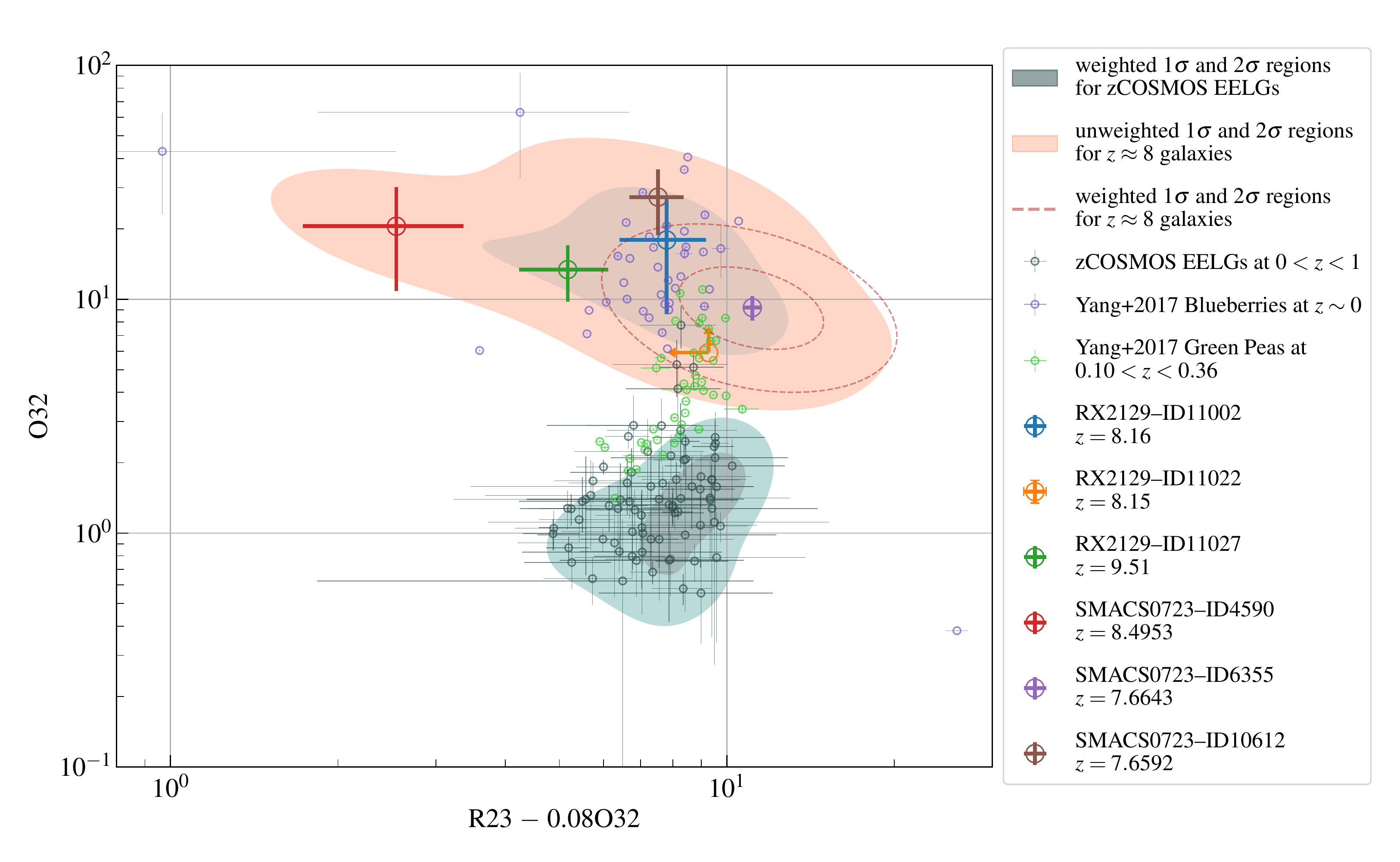}
    \caption{Strong line metallicity indicator comparison. This Figure shows O32 ($[$O{\footnotesize\;III}$]\lambda 5007$\AA/$[$O{\footnotesize\;II}$]\lambda \lambda 3727, 3729$\AA) plotted against R23$-$0.08$\times$O32 (R23$\times$H$\beta$ = $[$O{\footnotesize\;II}$]\lambda \lambda 3727, 3729$\AA + $[$O{\footnotesize\;III}$]\lambda 4959$\AA + $[$O{\footnotesize\;III}$]\lambda 5007$\AA) for the 6 galaxies at $z \approx 8$ with available NIRSpec strong emission line measurements (large colored data points). The shaded red region (dashed red line) show the unweighted (weighted) $1\sigma$ and $2\sigma$ confidence intervals of the $z \approx 8$ sample. For comparison we show the zCOSMOS extreme emission line galaxies (EELGs) at $z_{\textnormal{\scriptsize spec}} \lesssim 1$ (small dark-green data points) as well as their weighted $1\sigma$ and $2\sigma$ confidence intervals. We also show the blueberry (small purple data points) and green pea (small light-green data points) galaxies from \cite{yang+2017b, yang+2017g}. Both here and in Figure \ref{fig: OIII_to_OII}, blueberry galaxies (and green peas to a lesser degree) occupy a region similar to $z \approx 8$ emission line-detected galaxies, which indicates that they have remarkably similar strong emission line features and metallicities. However, as shown in Figure \ref{fig: MZ lowz}, the addition of the stellar mass parameter can distinguish these local analog candidates from $z \approx 8$ emission line galaxies in the context of the mass-metallicity relation.} 
    \label{fig: R23}
\end{figure*}

Figure \ref{fig: OIII_to_OII} shows the $[$O{\footnotesize\;III}$]\lambda 5007$\AA/H$\beta$ flux ratio plotted against the $[$O{\footnotesize\;II}$]\lambda \lambda 3727, 3729$\AA/H$\beta$ flux ratio for the galaxies in our $z \approx 8$ sample (large colored data points) for which these line intensity measurements are available; this only includes the six NIRSpec emission line-detected galaxies in RX2129 and SMACS0723. For the RX2129--ID11027 galaxy we use the line ratios as calculated in \cite{williams+2022}. For the SMACS0723 galaxies we use the line ratios from \citet{curti+2022}. We do not adopt any extinction correction for the $z \approx 8$ galaxies, consistent with the negligible extinction reported for the RX2129--ID11027 and SMACS0723 galaxies \citep[see][respectively]{williams+2022, curti+2022} as well as our photometry analysis of RX2129--ID11002 and RX2129--ID11022 in Section \ref{sec: photometry}. Figure \ref{fig: R23} shows the O32 plotted against (R23$-$0.08O32) for the galaxies in our $z \approx 8$ sample (large colored data points). 

The high $[$O{\footnotesize\;III}$]\lambda 5007$\AA/$[$O{\footnotesize\;II}$]\lambda \lambda 3727, 3729$\AA\ ratio of these $z \approx 8$ galaxies is typical of EELGs. In Figures \ref{fig: OIII_to_OII} and \ref{fig: R23}, for comparison we also include the EELGs from the $z \lesssim 1$ sample compiled by \cite{eelgs} from the zCOSMOS spectroscopic follow up survey \citep{zCOSMOS} of the COSMOS field \citep{COSMOS}. These authors report extinction-uncorrected emission line flux measurements as well as the reddening constant $c(\textnormal{H}\beta)$ derived from either the H$\alpha$/H$\beta$ or H$\gamma$/H$\beta$ ratios where available or spectral energy distribution fitting otherwise. We correct for extinction assuming a \cite{1989ApJ...345..245C} extinction law with $R_V = 3.1$. 

Young low-metallicity galaxies in the local Universe such as the green peas and blueberry galaxies \citep[see][]{2009MNRAS.399.1191C, yang+2017g, yang+2017b} are proposed as spectroscopic analogs of the high redshift galaxies. Both samples are selected as extreme emission line galaxies with systematically low metallicities at a fixed stellar mass. In Figure \ref{fig: OIII_to_OII} we also include the sample of green peas compiled in \cite{yang+2017g} as well as the blueberry galaxies from \cite{yang+2017b}. 

Figures \ref{fig: OIII_to_OII} and \ref{fig: R23} also show the $1\sigma$ and $2\sigma$ confidence intervals \citep[determined using the \texttt{seaborn} package;][]{seaborn} for the sample of $z \approx 8$ galaxies (dashed red contours) as well as the zCOSMOS EELGs (shaded green region), both calculated by weighting each entry in the sample by its $1\sigma$ line ratio uncertainties. We note that the confidence interval of the $z \approx 8$ sample is dominated by the SMACS0723--ID6355 galaxy which has much smaller line ratio uncertainties compared to the rest of this sample; therefore we also show the unweighted $1\sigma$ and $2\sigma$ confidence intervals for the $z \approx 8$ sample (shaded red region). The lack of overlap between the confidence interval regions of $z \approx 8$ galaxies and EELGs, even at the $2\sigma$ level, strongly suggests that these galaxies are drawn from intrinsically different populations.

Blueberry galaxies, and to a lesser degree green peas, show significant similarities to the $z \approx 8$ galaxies, with almost all the blueberry galaxies and $\sim$ half the green peas occupying the region within the $2\sigma$ confidence interval of $z \approx 8$ galaxies. This suggests that these galaxies have similar emission line features to those of the $z \approx 8$ galaxies, as well as similarly low metallicities. In Section \ref{sec: discussion low-z} we will further investigate this similarity in the context of the mass-metallicity relation, where the addition of the stellar mass parameter may potentially distinguish the blueberries (and green peas) from $z \approx 8$ emission line galaxies.

\subsection{Metallicity measurement} \label{sec: metallicity}

We measure the gas-phase metallicities\footnote{We use the terms ``gas-phase metallicity" and ``oxygen abundance" ($12 + \log\;$O/H) interchangeably throughout this work.} of RX2129--ID11002 and RX2129--ID11022 using the ``strong line" method empirical calibration from \cite{izotov+2019} 

\begin{equation}
    12 + \log\left (\frac{\textnormal{O}}{\textnormal{H}}\right) = 0.950 \log\left(\textnormal{R23} - 0.08 \times \textnormal{O32}\right) + 6.805\;.
\label{eq: metallicity}
\end{equation}


\noindent
This choice is motivated by the lack of significant detection of the $[$O{\footnotesize\;III}$]\lambda 4363$\AA\ emission line in both galaxies (which is required for the direct $T_e$ method) and is shown to measure accurate oxygen abundances for low-metallicity EELGs \citep[for a full discussion see][]{izotov+2019}. The measured metallicities of RX2129--ID11002 and RX2129--ID11022 are reported in Table \ref{table: line fluxes}. In this Table, we also report the total metallicity measurement uncertainty which includes both the statistical and systematic uncertainties; the former is the propagation of line flux measurement uncertainties and the latter is the 0.05 systematic uncertainty of the \cite{izotov+2019} ``strong line" metallicity calibration.

For the remaining galaxies in our $z \approx 8$ sample we use the metallicity measurements as reported in the literature. The metallicity of the RX2129--ID11027 galaxy was measured in \cite{williams+2022} using the ``strong line" method. The metallicities of the SMACS0723 galaxies were measured in \cite{curti+2022} and \cite{schaerer+2022} using the direct $T_e$ method, with significant discrepancies in the reported values. The main source of discrepancy seems to be the method used to reduce the NIRSpec data. We adopt the values reported by \cite{curti+2022} since the NIRSpec data used in this work had undergone extra reduction beyond the level 2 data products available on MAST (\href{https://archive.stsci.edu}{The Barbara A. Mikulski Archive for Space Telescopes}) through the NIRSpec GTO pipeline (NIRSpec/GTO collaboration, in preparation). Nevertheless, in Section \ref{sec: MZ} we investigate the effect of adopting the values reported in \cite{schaerer+2022}.

We note that the metallicities of these six NIRSpec emission line galaxies are not measured using the same method. This is forced by the lack of significant $[$O{\footnotesize\;III}$]\lambda 4363$\AA\ detections for the three RX2129 galaxies, preventing the application of the direct $T_e$ method. Alternatively, the strong line method can be used for the entire sample of NIRSpec emission line galaxies to achieve homogeneous metallicity measurements. This is presented in Appendix \ref{app: strong vs Te}, where the effect of adopting ``strong line" instead of ``direct $T_e$" metallicities for the SMACS0723 galaxies on the best-fit mass-metallicity relation is investigated in detail. In Appendix \ref{app: strong vs Te} we show that although the measured metallicities for two of the SMACS0723 galaxies differ slightly if the strong line method is used, the normalization and the slope of the best-fit mass-metallicity relation remain intact (see also Section \ref{sec: MZ}). Hence, throughout the rest of this work we adopt the direct $T_e$ metallicity measurements for the SMACS0723 galaxies since this method is believed to yield more accurate metallicity measurements.

\begin{figure*}
    \centering
    \includegraphics[width=18cm]{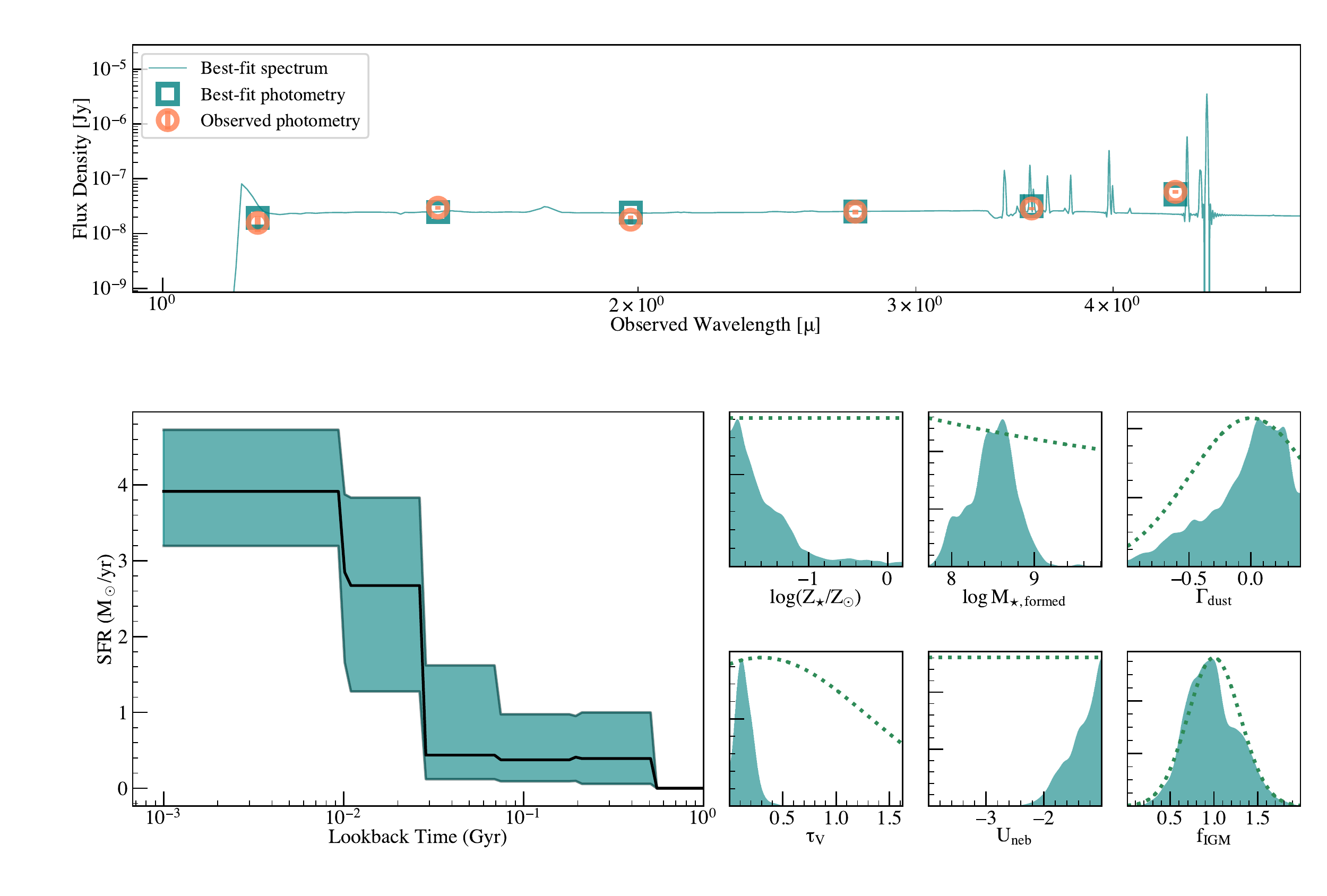}
    \caption{SED-fitting results for the RX2129--ID11002 galaxy. Top panel shows the observed (orange circles) and best-fit photometry (green squares) as well as the best-fit spectra (green line). The six smaller panels on the bottom right show the probability distribution functions (PDFs) of the stellar population synthesis parameters. The dotted lines show the assumed priors. The stellar mass here refers to the total stellar mass formed before correcting for the dead stars. Bottom left panel shows the star formation history (SFH) modelled nonparametrically with 5 temporal bins. All the parameters are corrected for lensing magnification.}
    \label{fig: 11002}
\end{figure*}

The metallicities of the pre-\jwst\ sample were measured in \cite{jones+2020}. These authors used a combination of the nebular $[$O{\footnotesize\;III}$]\lambda88\mu$m emission line and photometrically-measured star-formation rate (SFR) as a direct-method metallicity estimator (i.e. calibrated against the direct $T_e$ method). They report that their calibration yields $12 + \log\;$O/H values that are systematically offset by $+0.2$ from the direct $T_e$ method; we correct for this offset before adopting their measured metallicities. The offset-corrected values are reported in Table \ref{table: mass}. 

As demonstrated in \cite{jones+2020}, $[$O{\footnotesize\;III}$]\lambda88\mu$m-measured metallicities are in general not as accurate as metallicities measured through the rest-optical emission lines, namely the strong line and the direct $T_e$ methods. In particular, $[$O{\footnotesize\;III}$]\lambda88\mu$m-measured metallicities have a 0.2 dex $1\sigma$ scatter (after correction for the offset mentioned above) around the direct $T_e$ values. This is not a major source of concern for constraining the best-fit mass-metallicity relation since in this work both the statistical and the (large) systematic uncertainties of the $[$O{\footnotesize\;III}$]\lambda88\mu$m-measured metallicities are taken into account. This holds, if there is no significant systematic offset \citep[beyond the +0.2 dex value reported in][which has been corrected above]{jones+2020} between the $[$O{\footnotesize\;III}$]\lambda88\mu$m-measured metallicities and those measured through rest-optical emission lines.

We note that both the strong line and the $[$O{\footnotesize\;III}$]\lambda88\mu$m metallicity diagnostics are calibrated against the direct $T_e$ method at relatively low redshifts and in the local Universe, respectively. Future high S/N NIRSpec observations of high redshift emission line galaxies capable of detecting the $[$O{\footnotesize\;III}$]\lambda 4363$\AA\ line as well as the $[$O{\footnotesize\;III}$]\lambda \lambda 4959, 5007$\AA\ and $[$O{\footnotesize\;II}$]\lambda \lambda 3727, 3729$\AA\ doublets can be used to assess the strong line calibration against the direct $T_e$ method at these redshifts. This was investigated in Appendix \ref{app: strong vs Te} for the three SMACS0723 galaxies for which both metallicity diagnostics can be used; however, robust conclusions require larger samples. Similarly, NIRSpec follow-up observations of high redshift $[$O{\footnotesize\;III}$]\lambda88\mu$m emitters (e.g., see GO 1840) as well as ALMA follow-up observations of NIRSpec emission line galaxies will enable evaluating the calibration of $[$O{\footnotesize\;III}$]\lambda88\mu$m-measured metallicities against the rest-optical emission line methods at these redshifts, and in particular investigating if there is any systematic offset beyond the +0.2 dex value measured at the local Universe.

\section{Photometry analysis} \label{sec: photometry}

\begin{figure*}
    \centering
    \includegraphics[width=18cm]{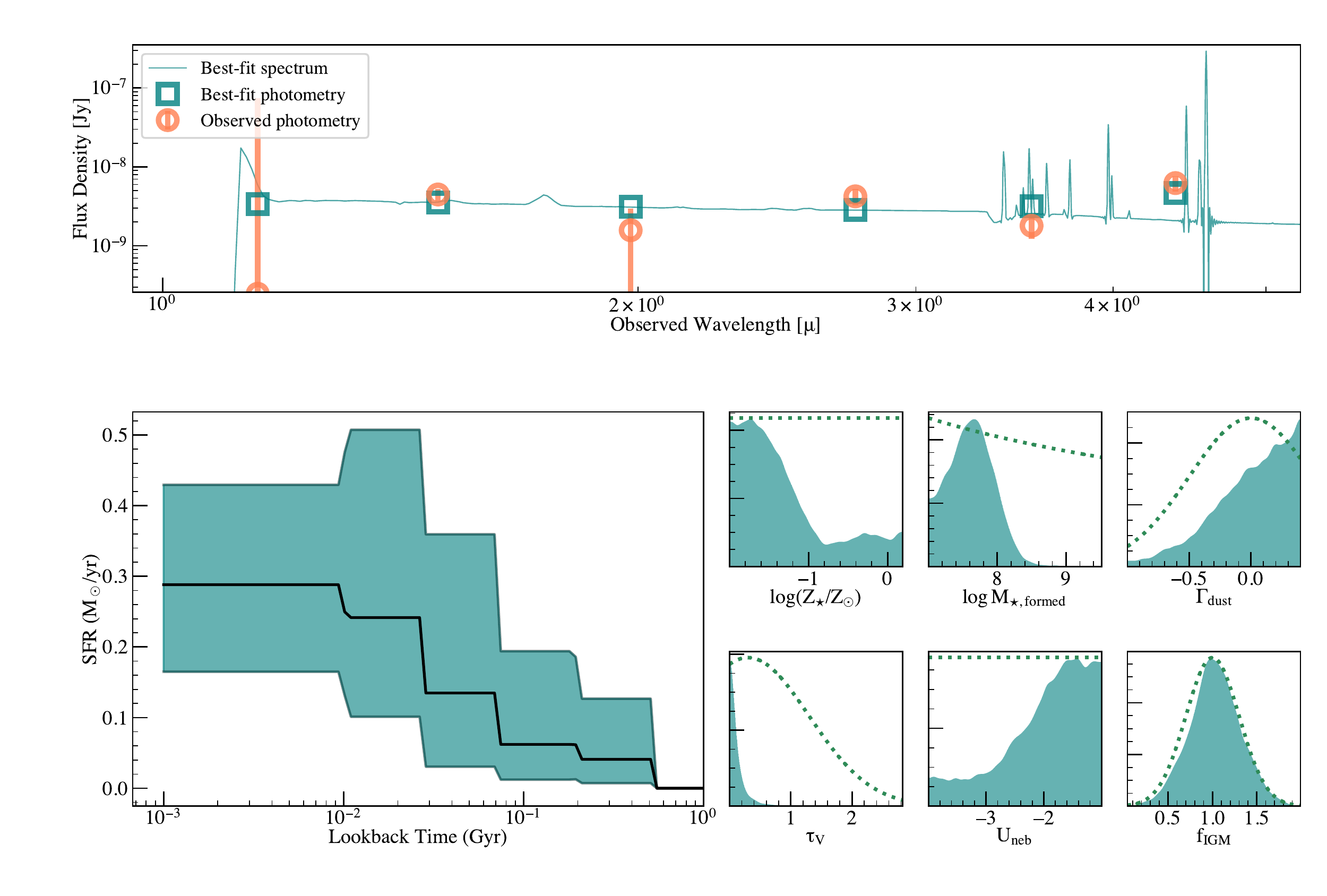}
    \caption{Same as Figure \ref{fig: 11002}, but showing the SED-fitting results for the RX2129--ID11022 galaxy.}
    \label{fig: 11022}
\end{figure*}

In this Section we use \texttt{prospector} \citep{prospector} to fit the available photometry of the galaxies in our $z \approx 8$ sample (see Section \ref{sec: sample}) and infer their stellar masses. \texttt{prospector} explores the posterior probability distributions of stellar population parameters to find the values that best fit the observed photometry. The spectra for each set of drawn stellar population parameters is derived from \texttt{FSPS} \citep[Flexible Stellar Population Synthesis;][]{FSPS1, FSPS2}, accessed through the \cite{2014zndo.....12157F} \texttt{python} bindings. Here we use the built-in \texttt{dynesty} sampler \citep{dynesty, 2022zndo...6609296K}, a \texttt{python}-based sampler adopting the dynamic nested sampling method developed by \cite{2019S&C....29..891H}. 

The \texttt{prospector} setup used in this work closely resembles that used in \cite{prospector} for fitting the measured photometry of GN-z11, the highest redshift ($z_{\textnormal{spec}} = 10.6$) spectroscopically-confirmed galaxy to date \citep{gnz11}\footnote{As of a few days after our initial submission, this galaxy no longer holds the record for the highest redshift spectroscopic confirmation; the \jwst\ Advanced Deep Extragalactic Survey program (JADES) has confirmed three galaxies at higher redshifts \citep{2022arXiv221204480R, 2022arXiv221204568C}.}. We fix the redshift to the spectroscopically measured value. The star-formation history (SFH) is modelled nonparametrically with 5 temporal bins. The last bin spans 0--10 Myr (lookback time). The remaining bins are evenly spaced in $\log(\textnormal{lookback time})$ up to the maximum allowed age of the galaxy as determined by its spectroscopic redshift and the earliest possible onset of star-formation, which we assume to be at $z = 35$. Our stellar population free parameters include the total formed stellar mass ($\textnormal{M}_{\star,\;\textnormal{{\scriptsize formed}}}$); the stellar metallicity ($\textnormal{Z}_{\star}$); the nebular metallicity ($\textnormal{Z}_{\textnormal{\scriptsize neb}}$); the nebular ionization parameter ($\textnormal{U}_{\textnormal{\scriptsize neb}}$, indicating the ratio of ionizing photons to the total hydrogen density); and the parameters controlling the dust attenuation and the intergalactic medium (IGM) attenuation.

We adopt the dust attenuation curve from \cite{2013ApJ...775L..16K} and include a diffuse dust component for the entire galaxy as well as a birth-cloud component for young stars. These two components are modelled with three free parameters: the diffuse dust optical depth at 5500\AA\ ($\tau_{\textnormal{\tiny V}}$); the ratio of the birth-cloud optical depth to the diffuse dust optical depth ($\textnormal{r}_{\textnormal{\scriptsize dust}}$); and the dust index ($\Gamma_{\textnormal{\scriptsize dust}}$) controlling the power-law slope of the attenuation curve. The intergalactic medium (IGM) attenuation is included as a free parameter because the rest-frame photometry at wavelengths below 1216\AA\ is affected by IGM attenuation; \texttt{prospector} adopts the redshift-dependent IGM attenuation model suggested by \cite{1995ApJ...441...18M}. The only free parameter is the IGM factor ($\textnormal{f}_{\textnormal{\scriptsize IGM}}$) which determines the normalization of \cite{1995ApJ...441...18M} model. 

The top panels in Figures \ref{fig: 11002} and \ref{fig: 11022} show the observed photometry and the best-fit (maximum a posteriori solution) spectra, respectively for RX2129--ID11002 and RX2129--ID11022. The 6 small panels at the bottom right of each Figure show the posterior probability distribution functions (PDFs) for a selection of free parameters. The dotted lines show the assumed priors. The bottom left panel shows the SFH. Similar Figures for the remaining 9 galaxies in our sample are available in Appendix \ref{app: prospector}. The stellar mass posterior PDFs in these Figures show the total formed stellar mass without subtracting the accumulated mass of dead stars; before further analysis, we correct this by applying the correction factors calculated by \texttt{prospector}. 
In Table \ref{table: mass} we report the best-fit (log probability-weighted 50th percentile of the posterior PDF) surviving stellar mass and its $1\sigma$ uncertainty (16th and 84th percentiles); however, throughout this work the full posterior PDFs are used whenever stellar mass measurements are needed.


Our best-fit stellar mass measurements for the pre-\jwst\ sample agree within $1\sigma$ with the lensing-corrected values used in \cite{jones+2020} which are adopted from \cite{rb2020}. These authors fit the photometry and the ALMA measurements of the $[$O{\footnotesize\;III}$]\lambda88\mu$m emission intensity and dust continuum with two-component SED models. The first component is a young starburst with strong nebular emission lines that contribute most of the flux in broad-band photometry and determines the $[$O{\footnotesize\;III}$]\lambda88\mu$m emission. The second component is a more mature stellar population that does not necessarily dominate the photometry but dominates the dust continuum detected in ALMA Band 7 and constitutes the majority of stellar mass. The authors show that unlike the models which fit the SFH with a single parameterized young component, these two-component models can simultaneously reproduce the dust continuum constraints and the broad-band photometry especially for MACS1149--JD1 and A2744--YD4. Based on a log-likelihood comparison between their two-component and single-component fits, the authors conclude that the two-component models provide superior fits to the data. This is further validated by our measurements which strongly rule out the values inferred by their single-component SED fits. We do not measure a significant systematic offset between our mass measurements and those of \cite{rb2020} \citep[$< 0.1$ dex if the B14--65666 galaxy, which has 1 dex errorbars in][is excluded]{rb2020}.

We note that there is a significant $\sim 1$ dex discrepancy between the stellar mass measurements of SMACS0723 galaxies reported in the literature \citep[see, e.g.,][]{curti+2022, schaerer+2022, tacchella+2022, carnall+2022}. The NIRCam photometry used in \cite{curti+2022} and \cite{schaerer+2022} were calibrated using the earlier versions of the calibration reference files (before the \texttt{jwst$\_$0989.map}), where flux calibration offsets as high as 0.2 mag exist between different filters \citep[see, e.g.,][]{2022RNAAS...6..191B}. This offset can potentially bias the inferred stellar mass, as is implied by the systematically higher values measured in both studies compared to \cite{carnall+2022} despite the relatively similar adopted SFH models. The photometry used in \cite{carnall+2022} has undergone extensive flux calibration as detailed in Appendix C of \cite{donnan+2022}, and is believed to be better calibrated compared to the calibrations achieved using the early NIRCam calibration reference files. 

The photometry used in our analysis is the same as that used in \cite{carnall+2022}. Nevertheless, the stellar mass measurements reported in \cite{carnall+2022} are systematically lower than our measurements by $> 0.7$ dex. These authors fit the photometry with a single-component parameterized (delayed exponential) SFH. As shown in \cite{rb2020} (see their Table 3), such SFH models do not account for the more mature stellar population which constitutes the overwhelming majority of stellar mass. \cite{rb2020} suggest that depending on the SFH such single-component models can underestimate the stellar mass by as much as 1.5 dex, well in line with the large offsets between our measurements and those of \cite{carnall+2022}. 

Our stellar mass measurements for SMACS0723--ID4590 and SMACS0723--ID10612 are consistent with the values inferred in \cite{tacchella+2022}, where both the NIRCam photometry and NIRSpec spectrum are simultaneously fitted to infer the stellar mass. This agreement is expected since \cite{tacchella+2022} used a \texttt{prospector} setup similar to our setup; in particular, these authors adopted a nonparametric SFH model. Compared to our measurements, \cite{tacchella+2022} measured a slightly higher ($< 0.5$ dex) stellar mass for SMACS0723--ID6355.


\section{Mass--metallicity relation} \label{sec: MZ}

In this Section we combine the metallicity measurements from Section \ref{sec: metallicity} and the stellar mass measurements from Section \ref{sec: photometry} (plotted as the colored data points in Figure \ref{fig: MZ lines}) to infer the mass--metallicity relation at $z \approx 8$. We use the method described in Appendix \ref{app: outlier} to fit the distribution of masses and metallicities with a linear relation of the form \citep[adopted from][]{ma+2016}

\begin{equation}
    12 + \log\left (\frac{\textnormal{O}}{\textnormal{H}}\right) - 9.0 = \gamma_{\textnormal{\scriptsize g}}\bigg[\log\left(\frac{M_{\star}}{M_{\odot}}\right) - 10\bigg] + \textnormal{Z}_{\textnormal{\scriptsize g,10}}\;,
\label{eq: forward}
\end{equation}

\noindent
where we search for the best-fit normalization $\textnormal{Z}_{\textnormal{\scriptsize g,10}}$ (gas-phase metallicity at $10^{10} M_{\odot}$ stellar mass) and slope $\gamma_{\textnormal{\scriptsize g}}$. The method adopted here (see Appendix \ref{app: outlier} for details) is best suited if it cannot be safely assumed that there are no outlier data point with severely underestimated uncertainties. As discussed in Section \ref{sec: metallicity}, this is likely the case for the measured metallicities. We find the best-fit values of $\gamma_{\textnormal{\scriptsize g}}$ = $0.24^{+0.13}_{-0.14}$ and $\textnormal{Z}_{\textnormal{\scriptsize g,10}}$ = $-1.08^{+0.19}_{-0.16}$. If the slope is fixed to the value found at lower redshifts \citep[$\gamma_{\textnormal{\scriptsize g}}$ = 0.30;][]{sanders+2021}, we find $\textnormal{Z}_{\textnormal{\scriptsize g,10}}$ = $-0.98^{+0.09}_{-0.15}$. These best-fit values are summarized in Table \ref{table: best-fit MZR}. 

Following the discussion in Section \ref{sec: metallicity}, we investigate if using the \cite{schaerer+2022} metallicity measurements for the SMACS0723 galaxies, instead of the values adopted in this work from \cite{curti+2022}, can significantly affect our results \citep[see][for a discussion on the different metallicity determinations of the three SMACS0723 galaxies]{2022ApJ...939L...3T}. Although this results in inferring a slightly shallower best-fit slope ($\gamma_{\textnormal{\scriptsize g}} = 0.21^{+0.15}_{-0.10}$), we do not report any meaningful change in its $1\sigma$ credible region. Similarly, the best-fit normalization does not change meaningfully.  

\begin{deluxetable}{lll}
\tablewidth{0pt}
\tablecaption{Best-fit slope ($\gamma_{\textnormal{\scriptsize g}}$) and normalization ($\textnormal{Z}_{\textnormal{\scriptsize g,10}}$) of the mass-metallicity relation (and their $1\sigma$ uncertainty) as defined by Equation \ref{eq: forward}.}
\label{table: best-fit MZR}
\tablehead{
\colhead{Model} &
\colhead{Slope} &
\colhead{Normalization} \\
\colhead{} &
\colhead{$\gamma_{\textnormal{\scriptsize g}}$} &
\colhead{$\textnormal{Z}_{\textnormal{\scriptsize g,10}}$}
}
\startdata
Equation \ref{eq: forward} with a fixed $\gamma_{\textnormal{\scriptsize g}}$ & 0.30 & $-0.98^{+0.09}_{-0.15}$ \\
Equation \ref{eq: forward} with a free $\gamma_{\textnormal{\scriptsize g}}$  & $0.24^{+0.13}_{-0.14}$ & $-1.08^{+0.19}_{-0.16}$ \\
\enddata
\end{deluxetable}

\begin{deluxetable*}{llllll}
\tablewidth{0pt}
\tablecaption{Measured stellar mass (lensing corrected) and metallicity for our $z \approx 8$ sample.}
\label{table: mass}
\tablehead{
\colhead{Galaxy} &
\colhead{$z_{\textnormal{\scriptsize spec}}$} &
\colhead{$\log(M_{\star}/M_{\odot})$} &
\colhead{$12 + \log(\textnormal{O/H})$} &
\colhead{Magnification}
 &
\colhead{References\tablenotemark{a}}
}
\startdata
RX2129--ID11002 & 8.160 & $8.49^{+0.24}_{-0.32}$ & $7.65^{+0.09}_{-0.09}$ & 2.23 & This work\\ 
RX2129--ID11022 & 8.150 & $7.52^{+0.33}_{-0.35}$ & $7.72$ \tablenotemark{b} & 3.29 & This work\\ 
RX2129--ID11027 & 9.510 & $7.74^{+0.23}_{-0.29}$ & $7.48^{+0.09}_{-0.09}$ & 19.60 & W22\\ 
SMACS0723--ID4590 & 8.498 & $8.00^{+0.36}_{-0.51}$ & $6.99^{+0.11}_{-0.11}$ & 10.09 & C23, D22\\ 
SMACS0723--ID6355 & 7.665 & $8.22^{+0.20}_{-0.18}$ & $8.24^{+0.07}_{-0.07}$ & 2.69 & C23, D22\\ 
SMACS0723--ID10612 & 7.663 & $8.40^{+0.15}_{-0.24}$ & $7.73^{+0.12}_{-0.12}$ & 1.58 & C23, D22\\ 
SXDF--NB1006--2 & 7.212 & $9.31^{+0.41}_{-0.47}$ & $7.36^{+0.41}_{-0.23}$ \tablenotemark{c} & 1.00 & I16\\ 
B14--65666 & 7.152 & $9.90^{+0.25}_{-0.33}$ & $7.94^{+0.21}_{-0.22}$ \tablenotemark{c} & 1.00 & H19, F16, B14\\ 
MACS0416--Y1 & 8.312 & $9.96^{+0.28}_{-0.23}$ & $8.03^{+0.21}_{-0.40}$ \tablenotemark{c} & 1.60 & T19, L15\\ 
A2744--YD4 & 8.382 & $10.03^{+0.40}_{-0.38}$ & $7.44^{+0.24}_{-0.26}$ \tablenotemark{c} & 1.50 & L17, L19, C16\\ 
MACS1149--JD1 & 9.110 & $9.31^{+0.19}_{-0.14}$ & $7.95^{+0.21}_{-0.21}$ \tablenotemark{c} & 11.50 & H18, L19, Z17\\ 
\enddata
\tablenotetext{a}{References for redshift and photometry. H18: \cite{H18}; L19: \cite{L19}; Z17: \cite{Z17}; L17: \cite{L17}; C16: \cite{C16}; T19: \cite{T19}; L15: \cite{L15}; C23: \cite{carnall+2022}; I16: \cite{I16}; H19: \cite{H19}; F16: \cite{F16}; B14: \cite{B14}; D22: \cite{donnan+2022}; W22: \cite{williams+2022} }\tablenotetext{b}{$1\sigma$ upper limit}
\tablenotetext{c}{Only the statistical uncertainty; an extra $\pm 0.2$ dex systematic uncertainty should be considered \citep[see][]{jones+2020}.}
\end{deluxetable*}

Moreover, we assess if defining the normalization at a stellar mass other than $M_{\star} = 10^{10}M_{\odot}$, which is the standard at lower redshifts, can affect our results. Equation \ref{eq: forward} explicitly assumes that the best-fit line passes from $\textnormal{Z}_{\textnormal{\scriptsize g,10}}$ at $M_{\star} = 10^{10}M_{\odot}$, which is $\sim 1$ dex higher than the average stellar mass of our $z \approx 8$ sample: $M_{\star} = 10^{8.8}M_{\odot}$. We modify this relation to instead infer the normalization at $M_{\star} = 10^{8.8}M_{\odot}$ (by replacing 10 with 8.8 in Equation \ref{eq: forward}). The results remain intact; we find a best-fit $\textnormal{Z}_{\textnormal{\scriptsize g,8.8}}$ = $-1.33^{+0.10}_{-0.09}$ (which corresponds to $\textnormal{Z}_{\textnormal{\scriptsize g,10}} = -1.06^{+0.10}_{-0.09}$) and $\gamma_{\textnormal{\scriptsize g}}$ = $0.21^{+0.15}_{-0.13}$, both in very good agreement with the results found adopting the standard normalization given by Equation \ref{eq: forward}. In the discussion of Section \ref{sec: discussion} and the rest of this work we adopt the best-fit mass-metallicity relation that was inferred above using the metallicities reported in \cite{curti+2022}, normalized at the standard stellar mass of $M_{\star} = 10^{10}M_{\odot}$ (see Table \ref{table: best-fit MZR}).

\begin{figure*}
    \centering
    \includegraphics[width=18cm]{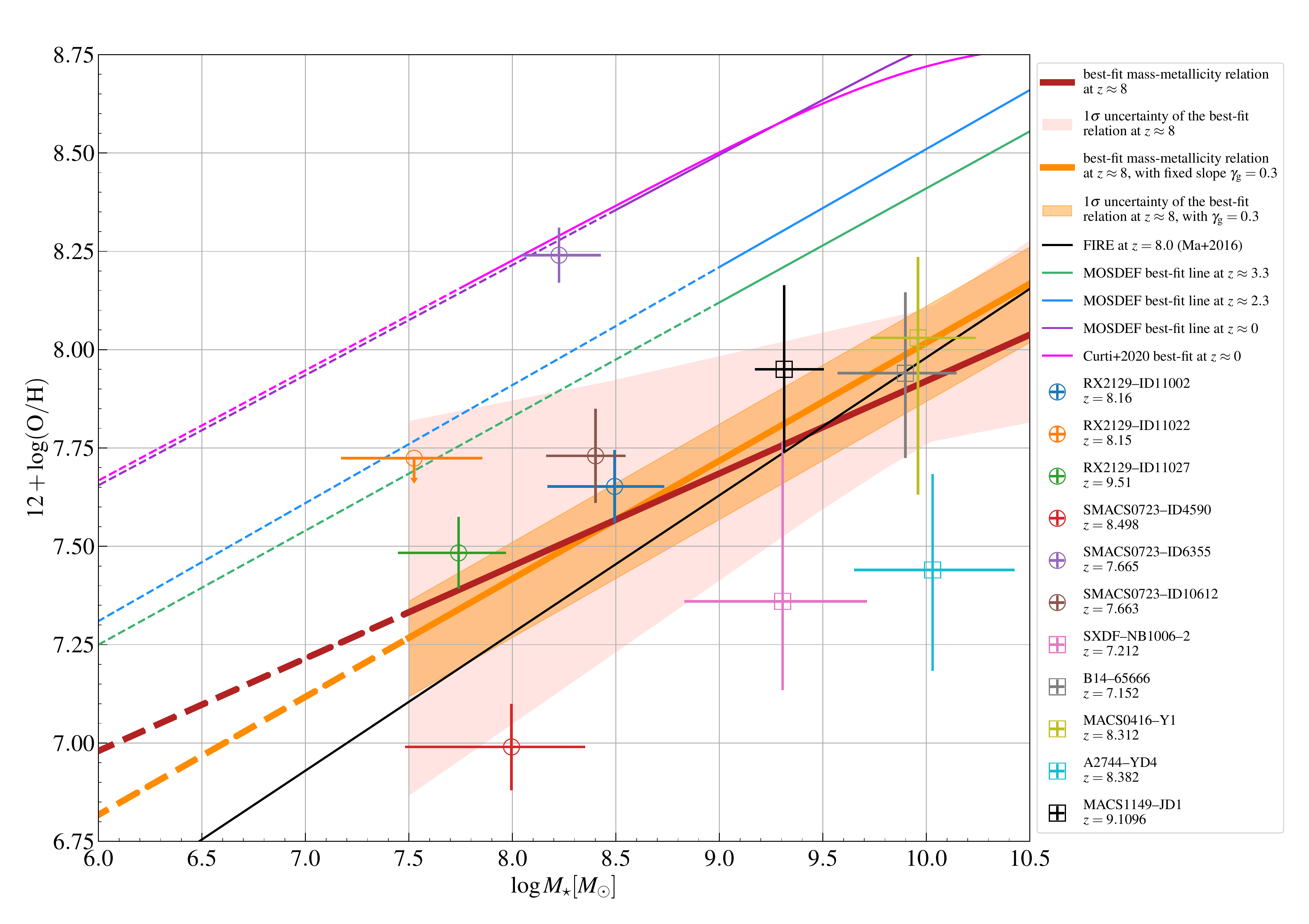}
    \caption{Mass--metallicity relation at $z \approx 8$. 
    Colored data points show the distribution of the measured masses and metallicities for the sources in our sample of $z \approx 8$ galaxies. The solid red line and the shaded pink region respectively show the best-fit $z \approx 8$ mass-metallicity relation and its $1\sigma$ uncertainty, if the mass-metallicity relation is fitted with a free slope. The solid orange line and the shaded orange region respectively show the best-fit $z \approx 8$ mass-metallicity relation and its $1\sigma$ uncertainty, as inferred by fixing the slope to the empirical value at lower redshifts, $\gamma_{\textnormal{\scriptsize g}} = 0.3$. The solid black line show the predicted mass-metallicity relation at $z = 8$ based on FIRE simulations \citep[see]{ma+2016}, showing remarkable agreement with our findings. For comparison, we also show the best-fit mass-metallicity relation at lower redshifts. The green, blue, and dark purple lines respectively show the best-fit mass-metallicity relation at $z \approx 3.3$, 2.3, and 0 inferred by \cite{sanders+2021} based on the data from MOSDEF survey. The light purple line shows the best-fit line at $z \approx 0$ from \cite{curti+2020b}. There is a significant $\sim 0.9$ dex evolution in the normalization of the mass-metallicity relation from $z \approx 8$ to the local Universe; on average galaxies are 8 times more metal enriched at $z \approx 0$ compared to $z \approx 8$. The evolution persists by 0.5 and 0.4 dex with respect to the average galaxies at $z \approx 2.3$ and $z \approx 3.3$, respectively. The dashed section of each solid line indicates extrapolation beyond the investigated stellar mass range of the corresponding study.}
    \label{fig: MZ lines}
\end{figure*}

We note that the rest-frame optical emission line ratios of $z \approx 8$ NIRSpec emission line-detected galaxies might be subject to contribution from AGN activity, which can potentially bias their measured metallicities. This speculation is particularly rooted in the increasing number of quasars detected at redshifts beyond $z = 6.5$ (\citealp[see, e.g.,][for the confirmed quasars]{2011Natur.474..616M, 2017ApJ...849...91M, 2018Natur.553..473B, 2018ApJ...869L...9W, 2019ApJ...884...30W, 2021ApJ...907L...1W, 2019ApJ...883..183M, 2019ApJ...872L...2M, 2019MNRAS.487.1874R, 2019AJ....157..236Y, 2020ApJ...897L..14Y}, \citealp[and][for a $z \approx 7.7$ candidate]{2022arXiv221210531F}). For instance, \cite{2022arXiv220807467B} argues that SMACS0723--ID6355 hosts a narrow-line AGN. Interestingly, this is the source with the most notable deviation from our best-fit mass-metallicity relation, as can be seen in Figure \ref{fig: MZ lines}. This is further emphasized by our method for finding the best-fit mass-metallicity relation, identifying the mass/metallicity measurements of this source as outliers ($66^{+21}_{-11}\%$ probability; see Appendix \ref{app: outlier} for more details). As detailed in Appendix \ref{app: outlier}, our method for fitting the mass-metallicity relation objectively prunes the outliers. Therefore, a few metallicity measurement outliers with severely underestimated uncertainties caused by AGNs are not expected to bias the inferred mass-metallicity relation. Significant biases can only be expected if a major fraction of our sources are affected by AGN activity. This is unlikely to be the case, especially for the $M_{\star} < 10^9 M_{\odot}$ galaxies (which constitute the majority of the discovered $z \approx 8$ NIRSpec emission line galaxies), a large fraction of which are not expected to host AGNs \citep[e.g., see][]{2019MNRAS.484.4413H} or host AGNs that are massive/active enough to significantly affect the rest-optical/UV spectrum \citep{2023MNRAS.521..241V}.

There is a clear separation in stellar mass between the NIRSpec emission line-detected galaxies ($\log M_{\star, \textnormal{mean}}/M_{\odot} = 8.3^{+0.1}_{-0.5}$; round data points in Figure \ref{fig: MZ lines}) and the pre-\jwst\ galaxies ($\log M_{\star, \textnormal{mean}}/M_{\odot} = 9.8^{+0.2}_{-0.4}$; squared data points in Figure \ref{fig: MZ lines}). The metallicities at the lower mass end are measured using the rest-optical emission lines and at the higher mass end using the far-infrared [\ion{O}{3}]$\lambda88\mu$m emission line. This dichotomy might affect the inferred normalization and slope of the mass-metallicity relation. We check if the inferred normalization is affected by separately fitting mass-metallicity relations to each mass range, with slopes fixed to the value found at lower redshifts ($\gamma_{\textnormal{\scriptsize g}}$ = 0.30). We find $\textnormal{Z}_{\textnormal{\scriptsize g,10}}$ = $-0.93^{+0.11}_{-0.16}$ for the six NIRSpec emission line-detected galaxies at the lower mass end, and $\textnormal{Z}_{\textnormal{\scriptsize g,10}}$ = $-1.09^{+0.18}_{-0.23}$ for the pre-\jwst\ galaxies at the higher mass end, well within $1\sigma$ uncertainties of one another. At this point, we cannot robustly evaluate the degree to which the inferred slope might be affected. This will become more clear with NIRSpec coverage of the higher mass end or with ALMA follow-up observations of NIRSpec emission line galaxies, providing more accurate calibrations of the [\ion{O}{3}]$\lambda88\mu$m metallicity diagnostic against the rest-optical diagnostics at $z \approx 8$ (see the discussion in Section \ref{sec: metallicity}).


\section{Discussion} \label{sec: discussion}

\begin{figure*}
    \centering
    \includegraphics[width=18cm]{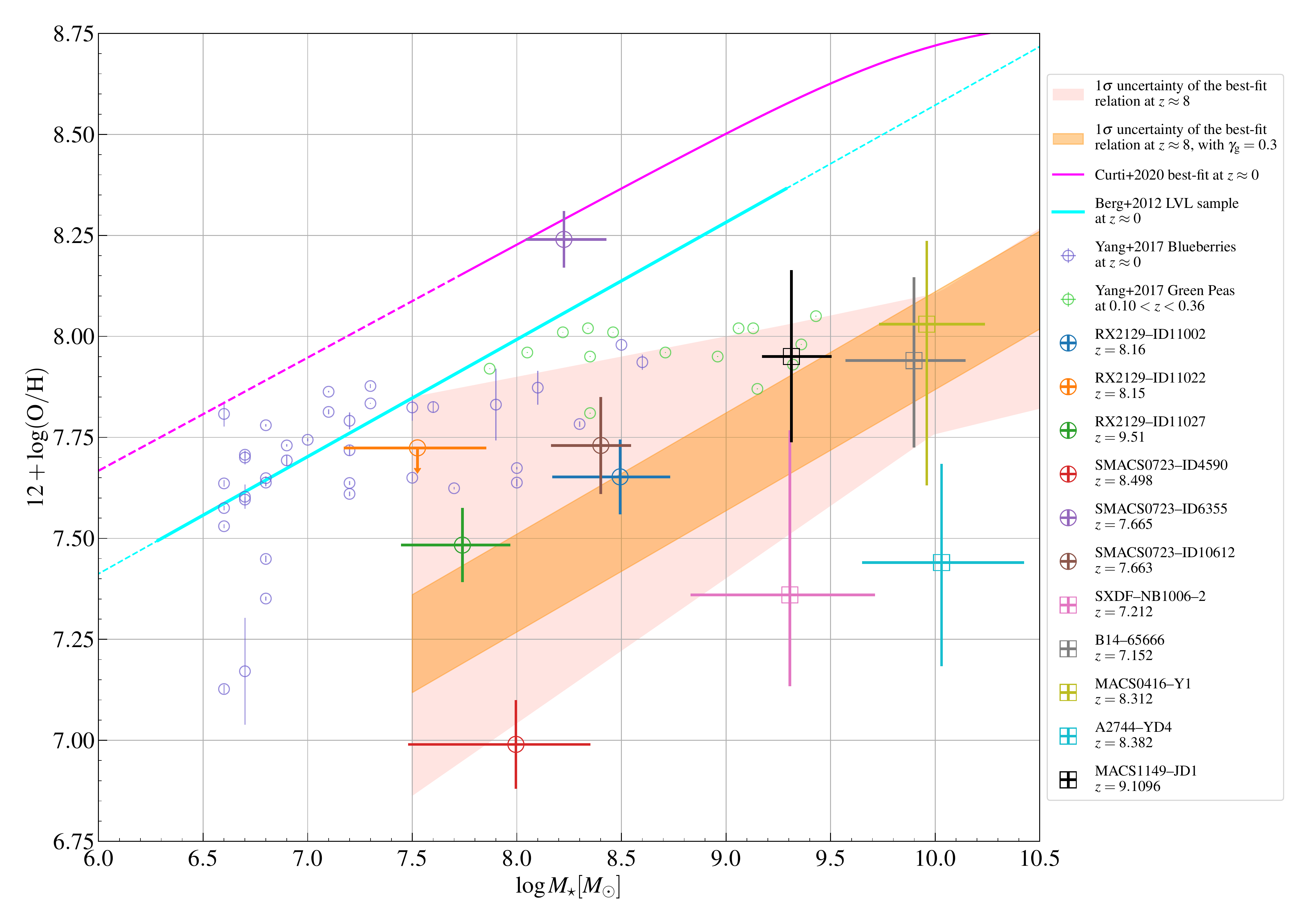}
    \caption{Mass-metallicity relation at $z \approx 8$ compared with the local analog candidates, the blueberry and green pea galaxies. The large colored data points show the measured mass and metallicity of the galaxies in our $z \approx 8$ sample; the solid orange (solid red) line shows the best-fit mass-metallicity relation with a fixed (free) slope and the shaded orange (shaded pink) region show its $1\sigma$ uncertainty region. Blueberry galaxies and green peas are shown with the small purple and light green data points, respectively. Only the green peas which lie within the $2\sigma$ confidence interval of the $z \approx 8$ galaxies in the $[$O{\footnotesize\;III}$]\lambda 5007$\AA/H$\beta$ vs $[$O{\footnotesize\;II}$]\lambda \lambda 3727, 3729$\AA/H$\beta$ and O32 vs R23$-$0.08$\times$O32 metallicity diagnostic plots (Figures \ref{fig: OIII_to_OII} and \ref{fig: R23}, respectively) are shown. We plot all the blueberry galaxies, because Figures \ref{fig: OIII_to_OII} and \ref{fig: R23} indicate that they have similar metallicities to the $z \approx 8$ emission line galaxies. This Figure shows that although blueberry galaxies and green peas have similar metallicities to the $z \approx 8$ galaxies, this degeneracy is broken down by considering the mass-metallicity relation; at a fixed stellar mass, green peas and blue berries are at systematically higher metallicity compare to $z \approx 8$ galaxies.}
    \label{fig: MZ lowz}
\end{figure*}

\subsection{Evolution of the mass-metallicity relation} \label{sec: MZR}

The best-fit $z \approx 8$ mass--metallicity relation as well as its $1\sigma$ uncertainty are shown in Figure \ref{fig: MZ lines} as the solid red line and the shaded pink region, respectively. First, we compare the normalization of the best-fit mass-metallicity relation at $z \approx 8$ with empirical constraints at lower redshifts, out to $z \approx 3.3$, based on the results of \cite{sanders+2021}. Our inferred normalization $\textnormal{Z}_{\textnormal{\scriptsize g,10}} = -1.08^{+0.19}_{-0.16}$ indicates a substantial $\sim 0.9$ dex redshift evolution, i.e., decrease in metallicity at a fixed stellar mass, with respect to the $z \approx 0$ measurements\footnote{This is measured at $10^{10} M_{\odot}$, and the evolution will be slightly smaller at lower stellar mass because of the different slopes of our best-fit line and that of \cite{sanders+2021}.}, where $\textnormal{Z}_{\textnormal{\scriptsize g,10}} = -0.18$ (\citealp[solid purple line in Figure \ref{fig: MZ lines}; see also,][]{ma+2016, maiolino+2019}; \citealp[the light pink line shows the best-fit $z \approx 0$ mass-metallicity relation from][which extends to lower stellar masses]{curti+2020b}). The inferred normalization at $z \approx 8$ also indicates significant redshift evolution with respect to $z \approx 2.3$ or $z \approx 3.3$ (solid blue and solid green lines, respectively), where $\textnormal{Z}_{\textnormal{\scriptsize g,10}} = -0.49$ and $-0.59$, respectively.

The slope $\gamma_{\textnormal{\scriptsize g}} = 0.24^{+0.13}_{-0.14}$ of the inferred mass-metallicity relation at $z \approx 8$ is slightly shallower than the slope measured at $z \approx 0$, 2.3, or 3.3 \citep[0.28, 0.30, 0.29, respectively; see][]{sanders+2021}, but they are consistent within $1\sigma$ uncertainties. To investigate this further, we adopted a fixed slope of $\gamma_{\textnormal{\scriptsize g}} = 0.3$ (consistent with the lower redshift observations) in Equation \ref{eq: forward} and repeated the analysis of Section \ref{sec: MZ} to find the best-fit normalization. This does not affect our inferred normalization $\textnormal{Z}_{\textnormal{\scriptsize g,10}} = -0.98^{+0.09}_{-0.15}$, still showing substantial evolution with respect to lower redshifts; this fit and its $1\sigma$ uncertainty are shown as the solid orange line and the shaded orange region in Figure \ref{fig: MZ lines}.

\cite{ma+2016} inferred the mass-metallicity relation out to $z = 6$ from the simulated galaxies in the FIRE project, showing good agreement with the available empirical measurements out to $z \approx 3.5$. Their inferred redshift evolution of $\textnormal{Z}_{\textnormal{\scriptsize g,10}}$ can be extrapolated beyond $z = 6$. They predict $\textnormal{Z}_{\textnormal{\scriptsize g,10}} = -0.98$ at $z = 6$ and $\textnormal{Z}_{\textnormal{\scriptsize g,10}} = -1.02$ at extrapolated $z = 8$, both in $1\sigma$ agreement with our measured normalization at $z \approx 8$. Nevertheless, our measurement mildly favors the extrapolated normalization at $z = 8$. Similar to the case of empirical measurements at lower redshifts, our measurement at $z \approx 8$ favors a shallower slope of the mass-metallicity relation compared to the \cite{ma+2016} predictions ($\gamma_{\textnormal{\scriptsize g}} = 0.35$), but the measurements are in agreement within $1\sigma$ uncertainties. The predicted $z = 8$ mass-metallicity relation of \cite{ma+2016} is shown as the solid black line in Figure \ref{fig: MZ lines}.

\subsection{Comparison with the low-redshift analogs} \label{sec: discussion low-z}

Figures \ref{fig: OIII_to_OII} and \ref{fig: R23} show that $z \approx 8$ emission line galaxies possess strong emission line features that are generally distinct from extreme emission lines galaxies (EELGs) at $z < 1$, but are similar to blueberry galaxies and, to some degree, green peas. Based on these plots we expect the $z \approx 8$ galaxies to have similar metallicities to blueberry galaxies and a subset of green peas. This might suggest blueberry galaxies (and green peas to a lesser degree) as local analogs to the $z \approx 8$ emission line galaxies. However, this analogy should be further considered in the context of mass-metallicity diagram.

Figure \ref{fig: MZ lowz} shows the distribution of the measured stellar masses and metallicities of our sample of $z \approx 8$ galaxies (large colored data points), as well as the $1\sigma$ uncertainty of their distribution around the best-fit mass-metallicity relation (shaded pink and shaded orange regions, respectively for a free and fixed slope, see Section \ref{sec: MZR}). The small purple data points show the distribution of blueberry galaxies, all of which show similar strong emission line features to those of $z \approx 8$ emission line galaxies based on Figures \ref{fig: OIII_to_OII} and \ref{fig: R23}. We also show a sub-sample of green peas (small green data points), consisting of the green peas that lie within the $2\sigma$ credible interval of the $z \approx 8$ galaxies in Figures \ref{fig: OIII_to_OII} and \ref{fig: R23}. Both the blueberry galaxies and green peas have metallicities similar to the $z \approx 8$ galaxies, as expected. However, at a fixed stellar mass, the $z \approx 8$ galaxies populate a region with lower metallicity compared to green peas and blueberry galaxies, and hence stand out in the mass-metallicity diagram. 

In Figure \ref{fig: MZ lowz}, we also show the \cite{berg+2012} $z \approx 0$ mass-metallicity relation for dwarf galaxies in the \textit{Spitzer} Local Volume Legacy (LVL) survey \citep{2009ApJ...703..517D}. The \cite{berg+2012} mass-metallicity relation is also consistent with the distribution of low metallicity galaxies from the Dark Energy Survey \citep{2022arXiv221102094L}. We converted the stellar masses reported in \cite{berg+2012} from Salpeter IMF to Chabrier IMF. This is not a representative sample of galaxies at $z \approx 0$ but rather biased toward lower metallicities \citep[e.g., see the discussion in][]{curti+2020b}. Nevertheless, it is interesting to note that the \cite{berg+2012} mass-metallicity relation coincides with the location of blueberries in this diagram. Despite the lower normalisation of this relation compared to the representative mass-metallicity relation at $z \approx 0$ (the \cite{curti+2020b} relation is shown as the solid pink line in Figure \ref{fig: MZ lowz}), it is systematically at higher normalization than that of the $z\approx8$ galaxies. 

There are remarkable strong emission line and metallicity similarities between the $z \approx 8$ emission line galaxies and extremely low metallicity local analog candidates, especially blueberry galaxies. However, the persistent systematically lower metallicities of $z \approx 8$ galaxies at a fixed stellar mass with respect to the local analog candidates suggests potential differences in their star formation, feedback, and enrichment histories. In \cite{2023arXiv230706336L}, we further investigate the metallicity deficiency of high redshift galaxies in the context of the fundamental metallicity relation, showing that the ultraviolet compactness can be used as a tracer of lowest metallicity galaxies.

\section{Conclusion} \label{sec: conclusion}

We present the \jwst\ NIRCam photometry and NIRSpec spectra of two galaxies at $z \approx 8.15$ detected in the field towards the lensing cluster RX\,J2129.4$+$0005. We used these galaxies as well as 9 other galaxies at $7 < z < 9$ from the literature to compile a sample of 11 galaxies at $z \approx 8$ with stellar mass and gas-phase metallicity measurements. Using this sample we establish the mass-metallicity relation at $z \approx 8$.

The normalization of the mass-metallicity relation has evolved significantly from $z \approx 8$ to the local Universe; metallicity at a fixed stellar mass has increased significantly from $z \approx 8$ to $z \approx 0$. We compared our results with the mass-metallicity relation inferred by \cite{sanders+2021} at $z \approx 0$. The normalization of our best-fit mass-metallicity relation at $z \approx 8$ is $\sim 0.9$ dex lower than the normalization at $z \approx 0$; galaxies are on average 8 times less metal enriched at $z \approx 8$ compared to the local Universe. Furthermore, the evolution persists by $\sim 0.5$ dex and $\sim 0.4$ dex respectively, compared to the $z \approx 2.3$, 3.3 empirical results of \cite{sanders+2021}. The galaxies observed at $z \approx 8$ are on average half as enriched as the galaxies at $z \approx 3.3$, the highest redshift up to which the mass-metallicity relation has been probed prior to \jwst. This implies a remarkably rapid enrichment epoch in the early Universe, when in less than $3.5\%$ of the lifetime of an average galaxy ($ < 450$ Myr at $z \approx 8$, assuming the galaxy starts forming at $z = 20$) almost $10\%$ of its enrichment has already happened. 
    
In general, our results agree well with the evolution of the mass-metallicity relation as predicted by the FIRE simulations \citep{ma+2016}. Our measured normalization of the mass-metallicity relation at $z \approx 8$ agrees within a few $0.01$ dex with \cite{ma+2016} predictions, well below the statistical uncertainty of our measurement. 
    
We tested the particular case where we did not fix the slope of the best-fit mass-metallicity relation to the slope suggested based on simulations or lower redshift observations. In this case, our inferred slope ($\gamma_{\textnormal{\scriptsize g}} = 0.24$) is slightly shallower than the measured slope at lower redshift ($\gamma_{\textnormal{\scriptsize g}} = 0.3$) or the slope predicted by simulation ($\gamma_{\textnormal{\scriptsize g}} = 0.35$). However, we cannot rule out these slopes, since they are within the $1\sigma$ uncertainty of our measurement. Compiling larger samples of $z \approx 8$ galaxies will address this further. 
    
We compared the $z \approx 8$ galaxies with potential analogs in the low redshift Universe, based on the $[$O{\footnotesize\;III}$]\lambda 5007$\AA/H$\beta$ vs $[$O{\footnotesize\;II}$]\lambda \lambda 3727, 3729$\AA/H$\beta$ and the O32 vs (R23-0.08O32) diagnostic plots. We find that galaxies detected at $z \approx 8$ have emission line features that are distinct from extreme emission line galaxies at $z \approx 0-1$, and have systematically lower metallicities. However, there seems to be remarkable similarities in the emission line features of the blueberry galaxies (and to some degree green peas) and the $z \approx 8$ emission line galaxies. We investigated this further in the context of the mass-metallicity diagram: at a fixed stellar mass, the $z \approx 8$ galaxies have systematically lower metallicities compared to blueberry galaxies and therefore stand out in the mass--metallicity diagram.

\begin{acknowledgments}
We thank the anonymous referee whose comments helped improve the robustness of our analysis and conclusions.
We thank Gabe Brammer for his identification of the high-redshift galaxy targets we present during the mask design process, and his contributions to their analysis. We also appreciate Program Coordinator Patricia Royle, and Program Scientists Armin Rest, Diane Karakala, and Patrick Ogle for their efforts with short turnaround that made the follow-up observations a success. D.L. and J.H. were supported by a VILLUM FONDEN Investigator grant (project number 16599).
P.L.K. is supported by NSF grant AST-1908823 and anticipated funding from {\it JWST} DD-2767. 
A.Z. acknowledges support by Grant No. 2020750 from the United States-Israel Binational Science Foundation (BSF) and Grant No.\ 2109066 from the United States National Science Foundation (NSF), and by the Ministry of Science \& Technology, Israel.
J.M.D. acknowledges the support of projects PGC2018-101814-B-100 and MDM-2017-0765. 
A.V.F. is grateful for financial assistance from the Christopher R. Redlich Fund and numerous individual donors.  
The UCSC team is supported in part by NSF grant AST--1815935, the Gordon \& Betty Moore Foundation, and by a fellowship from the David and Lucile Packard Foundation to R.J.F.
M.O. acknowledges support by JSPS KAKENHI grants JP20H00181, JP20H05856, JP22H01260, and JP22K21349.
I.P.-F. and F.P. acknowledge support from the Spanish State Research Agency (AEI) under grant number PID2019-105552RB-C43.
J.P. was supported by HST program GO-16264 through the Space Telescope Science Institute, which is operated by the Association of Universities for Research in Astronomy, Inc.\ for NASA, under contract NAS5-26555.

The \jwst\ NIRCam and NIRSpec data presented/used in this paper were obtained from the Mikulski Archive for Space Telescopes (MAST) at the Space Telescope Science Institute. The specific observations analyzed can be accessed via \dataset[DOI]{https://doi.org/10.17909/hnrr-sc32}.
\end{acknowledgments}

\software{
dynesty \citep{dynesty},
EAZY \citep{brammervandokkumcoppi08},
EMCEE \citep{emcee},
FSPS \citep{FSPS1, FSPS2}, 
glafic \citep{oguri+2010, oguri+2021},
prospector \citep{prospector}, 
pPXF \citep{2004PASP..116..138C, 2017MNRAS.466..798C, 2022arXiv220814974C}, 
seaborn \citep{seaborn},
Zitrin-parametric code \citep{zitrin+2015}
}

\vspace{20mm}
\bibliography{main}
\bibliographystyle{aasjournal}







\appendix

\section{Fitting the mass-metallicity relation} \label{app: MZ relation}

In this Section we describe the method used in Section \ref{sec: MZ} to find the best-fit mass-metallicity relation. First we describe the method used when it can be safely assumed that there are no outlier data points with severely underestimated uncertainties (\ref{app: no outlier}); later in this Appendix we describe the method used when this assumption is prohibited (\ref{app: outlier}).

\subsection{Assuming that there are no outlier data points} \label{app: no outlier}

In order to find the best linear fit of the form given in Equation \ref{eq: forward} we explore the parameter space of $\gamma_{\textnormal{\scriptsize g}}$ and $\textnormal{Z}_{\textnormal{\scriptsize g,10}}$ to find their posterior PDFs that best describe the measured masses and metallicities as well as their uncertainties. For this purpose we use the \texttt{EMCEE} package \citep{emcee}, a \texttt{python} implementation of the affine-invariant ensemble sampler \citep{2010CAMCS...5...65G} for Markov chain Monte Carlo (MCMC). At each MCMC step, we randomly draw the stellar mass of each ($i$'th) source $M_{\star, i}$ from its full PDF (see Section \ref{sec: photometry}) and search for the $\gamma_{\textnormal{\scriptsize g}}$ and $\textnormal{Z}_{\textnormal{\scriptsize g,10}}$ values which maximize the probability defined as

\begin{equation}
\begin{split}
    \ln\mathcal{L}_{\textnormal{normal}} = \ln p\bigg(\big\{12+\log(\textnormal{O/H})_{i, \textnormal{\scriptsize truth}}\big\}_{i=1}^N\bigg| \big\{M_{\star, i}\big\}_{i=1}^N, \gamma_{\textnormal{\scriptsize g}}, \textnormal{Z}_{\textnormal{\scriptsize g,10}}\bigg) = \\
    \sum_{i=1}^{N} \ln\left( \frac{1}{\sigma_i \sqrt{2\pi}} \right) - 0.5 \frac{\big(\log(\textnormal{O/H})_{i, \textnormal{\scriptsize truth}} - \log(\textnormal{O/H})_{i, \textnormal{\scriptsize model}}\big)^2}{\sigma_i^2}\;,
\end{split}
\end{equation}

\noindent
where the sum is over the entire sample of galaxies; $\log(\textnormal{O/H})_{i, \textnormal{\scriptsize truth}}$ is the measured metallicity of each source (from Section \ref{sec: metallicity}); $\log(\textnormal{O/H})_{i, \textnormal{\scriptsize model}}$ is calculated by inserting the drawn stellar mass, $\gamma_{\textnormal{\scriptsize g}}$, and $\textnormal{Z}_{\textnormal{\scriptsize g,10}}$ in Equation \ref{eq: forward}; and

\begin{equation}
    \sigma_i =     
    \begin{cases}
      \textnormal{positive uncertainty of } \log(\textnormal{O/H})_{i, \textnormal{\scriptsize truth}}, 
      \;\;\;\text{if}\ \log(\textnormal{O/H})_{i, \textnormal{\scriptsize model}} \ge \log(\textnormal{O/H})_{i, \textnormal{\scriptsize truth}} \\
      \textnormal{negative uncertainty of } \log(\textnormal{O/H})_{i, \textnormal{\scriptsize truth}},
      \;\;\;\text{if}\ \log(\textnormal{O/H})_{i, \textnormal{\scriptsize model}} < \log(\textnormal{O/H})_{i, \textnormal{\scriptsize truth}}
    \end{cases}
    .
\label{eq: sigma}
\end{equation}

\noindent
Following this approach we use the full posterior PDFs of the mass measurements and assume that the posterior PDFs of metallicity measurements are described by split normal distributions. It is easy to see that in the limit where the mass PDFs approach delta functions centred on the maximum a posteriori value, our defined probability function approaches the familiar case of log likelihood maximization for negligible uncertainties on mass measurements. 

\subsection{Assuming that there might be outlier data points} \label{app: outlier}

The above approach is not robust when there are outlier measurements with severely underestimated uncertainties. As discussed in Section \ref{sec: metallicity}, this is most likely the case for the metallicity measurements of the $z \approx 8$ sample. In order to objectively prune the outliers, we modify our probability function following the method suggested by \cite{hogg+2010}. This corresponds to adding $3 + N$ new free parameters including $q_i$ (one per data point) each of which is zero if the corresponding data point is believed to be an outlier and is one if the corresponding data point is believed to not be an outlier; $P_b$ which is the prior probability that any individual data point is an outlier; and $Y_b$ and $V_b$ which determine the mean and variance of the outliers. 
The modified probability function takes the form 

\begin{equation}
\begin{split}
    \ln\mathcal{L}_{\textnormal{prune}} = \sum_{i=1}^{N} \bigg[ \ln( \frac{1}{\sigma_i \sqrt{2\pi}} ) - 0.5 \frac{\big(\log(\textnormal{O/H})_{i, \textnormal{\scriptsize truth}} - \log(\textnormal{O/H})_{i, \textnormal{\scriptsize model}}\big)^2}{\sigma_i^2} \bigg] \times q_i \; + 
    \\
    \sum_{i=1}^N \bigg[\ln(\frac{1}{\sqrt{2\pi (V_b + \sigma_i^2})}) - 0.5 \frac{(12 + \log(\textnormal{O/H})_{i, \textnormal{\scriptsize truth}} - Y_b)^2}{V_b + \sigma_i^2} \bigg] \times (1-q_i) \;,
\end{split}
\end{equation}

\noindent
where $\sigma_i$ is given by Equation \ref{eq: sigma}. In order to penalize data rejection, we include a prior probability on $q_i$, given by Equation 14 in \cite{hogg+2010}

\begin{equation}
    \ln p(\big\{q_i\big\}_{i=1}^N \big| P_b) = \sum_{i=1}^{N} q_i \ln(1-P_b) + (1-q_i) \ln P_b \;.
\end{equation}

\noindent
We adopt a flat prior in the range $[0,1]$ on $P_b$, and flat priors in ranges $[6,9]$ and $[0,6]$ on $Y_b$ and $V_b$ motivated by the range of $12 + \log(\textnormal{O/H})_{i, \textnormal{\scriptsize truth}}$. We marginalize over the nuisance parameters, $\{q_i\}_{i=1}^N, P_b, Y_b, V_b$, to report the best-fit $\gamma_{\textnormal{\scriptsize g}}$ and $\textnormal{Z}_{\textnormal{\scriptsize g,10}}$. The strength of this method is that, apart from the parameters of interest, it also constraints the posterior PDF of a given data point being an outlier ($q_i$).

\section{Strong line method metallicity measurements for the SMACS0723 galaxies} \label{app: strong vs Te}

As mentioned in Section \ref{sec: metallicity}, the methods used to measure the metallicities of our sample of NIRSpec emission line-detected galaxies are not homogeneous: the direct $T_e$ method was used to measure the metallicities of the SMACS0723 galaxies, while the strong line method was used for the RX2129 galaxies. Here we measure the metallicities of the SMACS0723 galaxies using the strong line method in order to build a sample with more homogeneously measured metallicities: Six NIRSpec emission line galaxies with strong line method metallicity measurements and five pre-\jwst\ galaxies with ALMA [\ion{O}{3}]$\lambda88\mu$m emission line metallicity measurements. We use this sample to repeat the analysis of Section \ref{sec: MZ} and find the best-fit mass-metallicity relation for the multiple setups considered in this work: adopting Equation \ref{eq: forward} with a fixed slope ($\gamma_{\textnormal{\scriptsize g}} = 0.3$); adopting Equation \ref{eq: forward} with a free slope; and adopting Equation \ref{eq: forward} with a free slope with the normalization defined at $10^{8.8} M_{\odot}$ (the average stellar mass of our sample) instead of $10^{10} M_{\odot}$ (see Section \ref{sec: MZ} for more details on the motivation for this setup).

\begin{figure*}[b]
    \centering
    \includegraphics[width=14cm]{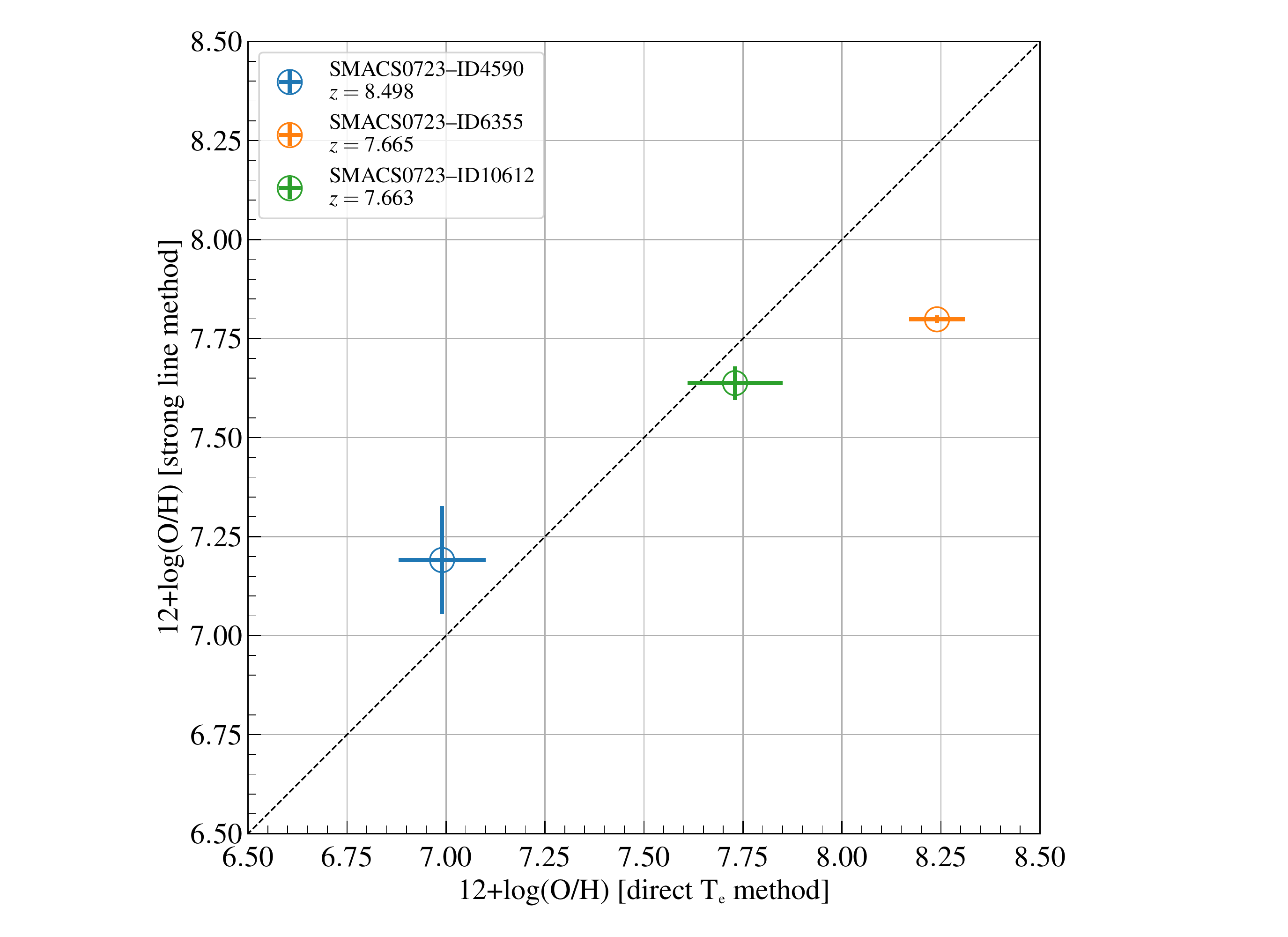}
    \caption{``Strong line" metallicity measurements plotted against the ``direct $T_e$" metallicity measurements \citep[from][]{curti+2022} for the SMACS0723 galaxies. Although metallicities for individual galaxies differ slightly depending on the adopted metallicity measurement method, we confirm that the overall effect on the best-fit mass-metallicity relation is negligible.}
    \label{fig: strong vs Te}
\end{figure*}

Using the strong line method, we measure $12 + \log(\textnormal{O/H})$ = $7.19 \pm 0.14$, $7.80 \pm 0.010$, and $7.64 \pm 0.04$ for SMACS0723--ID4590, SMACS0723--ID6355, and SMACS0723--ID10612, respectively. We compare the strong line measurements with the direct $T_e$ measurements in Figure \ref{fig: strong vs Te}. SMACS0723--ID6355 is the galaxy where the strong line method and the direct $T_e$ method are most in disagreement. Interestingly this is the galaxy that is most offset from our best-fit mass-metallicity relation (see Figure \ref{fig: MZ lines}), most confidently identified as an outlier metallicity measurement with severely underestimated uncertainties by the algorithm we used to fit the mass-metallicity relation (see Appendix \ref{app: outlier}), and also is suggested to likely host an AGN by \cite{2022arXiv220807467B}.

Using these new metallicity measurements and following the method used in Section \ref{sec: MZ} (see also Appendix \ref{app: outlier}), we measure a best-fit $\textnormal{Z}_{\textnormal{\scriptsize g,10}}$ (normalization) = $-1.03^{+0.10}_{-0.10}$ for the mass-metallicity relation if the slope is fixed to $\gamma_{\textnormal{\scriptsize g}} = 0.3$; this is in very good agreement with the normalization measured for a similar setup in Section \ref{sec: MZ} where the direct $T_e$ metallicities for the SMACS0723 galaxies are adopted. Similarly, we measure $\textnormal{Z}_{\textnormal{\scriptsize g,10}}$ = $-1.05^{+0.13}_{-0.15}$ and $\gamma_{\textnormal{\scriptsize g}}$ = $0.24^{+0.11}_{-0.11}$ if the slope is free, and $\textnormal{Z}_{\textnormal{\scriptsize g,8.8}}$ = $-1.31^{+0.09}_{-0.08}$ and $\gamma_{\textnormal{\scriptsize g}}$ = $0.23^{+0.13}_{-0.15}$ if the slope is free and the normalization is measured at $10^{8.8} M_{\odot}$, both in very good agreement with the values reported in Section \ref{sec: MZ} where the direct $T_e$ metallicities for the SMACS0723 galaxies are adopted. Hence, although the metallicities for individual SMACS0723 galaxies differ slightly if the strong line method is used instead of the direct $T_e$ method, the overall effect on the best-fit mass-metallicity relation seems negligible.

\section{Best-fit stellar populations to our sample of \titlelowercase{$z \approx 8$} galaxies} \label{app: prospector}

In this Appendix we show the best-fit photometry and spectrum, the posterior PDFs of the stellar populations, and the SFHs for the remaining 9 sources in our sample of $z \approx 8$ galaxies; the same plots for the RX2129--ID11002 and RX2129--ID11022 galaxies were shown in Section \ref{sec: photometry} (Figures \ref{fig: 11002} and \ref{fig: 11022}).

\begin{figure*}[hb]
    \centering
    \includegraphics[width=18cm]{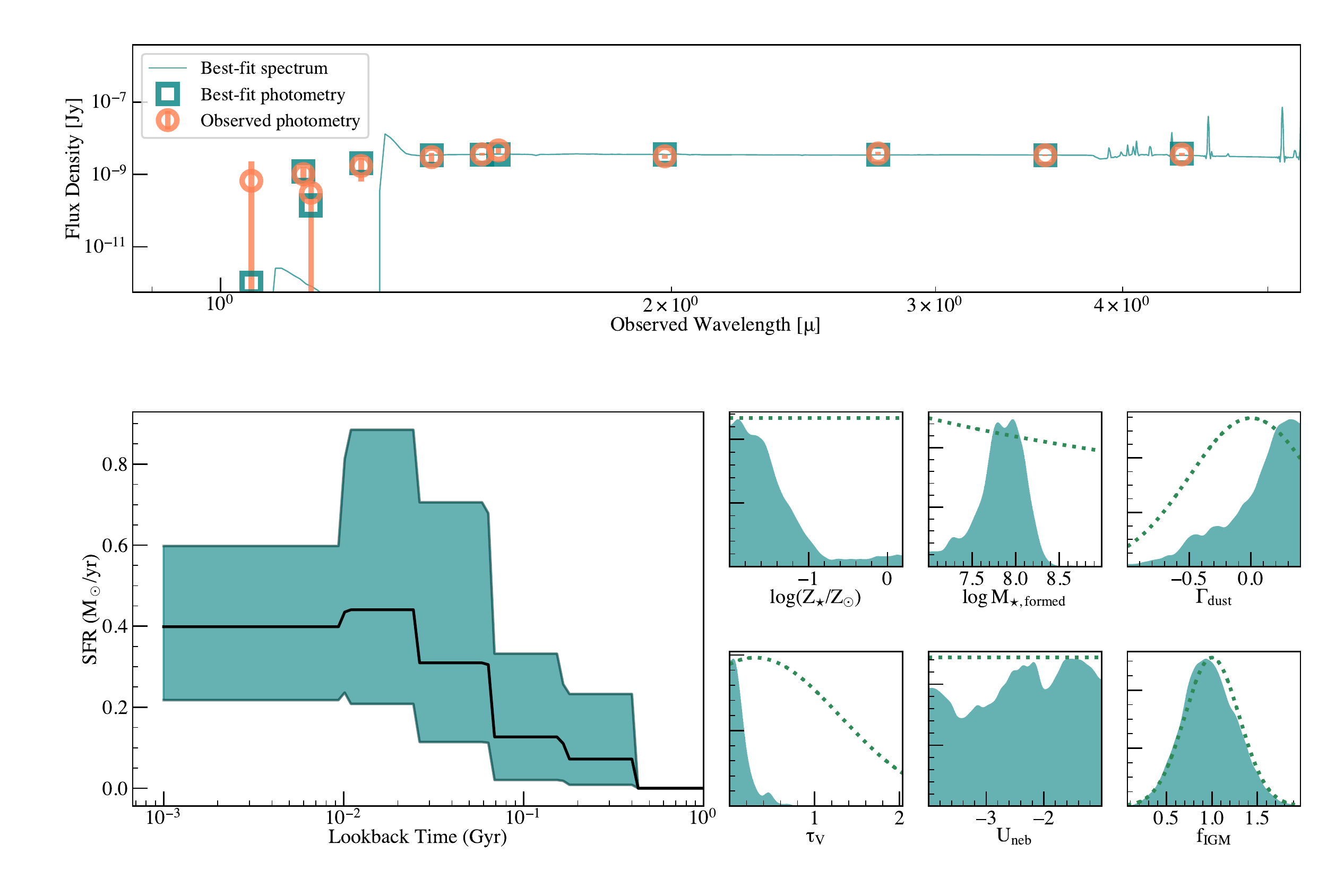}
    \caption{SED-fitting results for the RX2129--ID11027 galaxy.}
    \label{fig: 11027}
\end{figure*}

\begin{figure*}
    \centering
    \includegraphics[width=18cm]{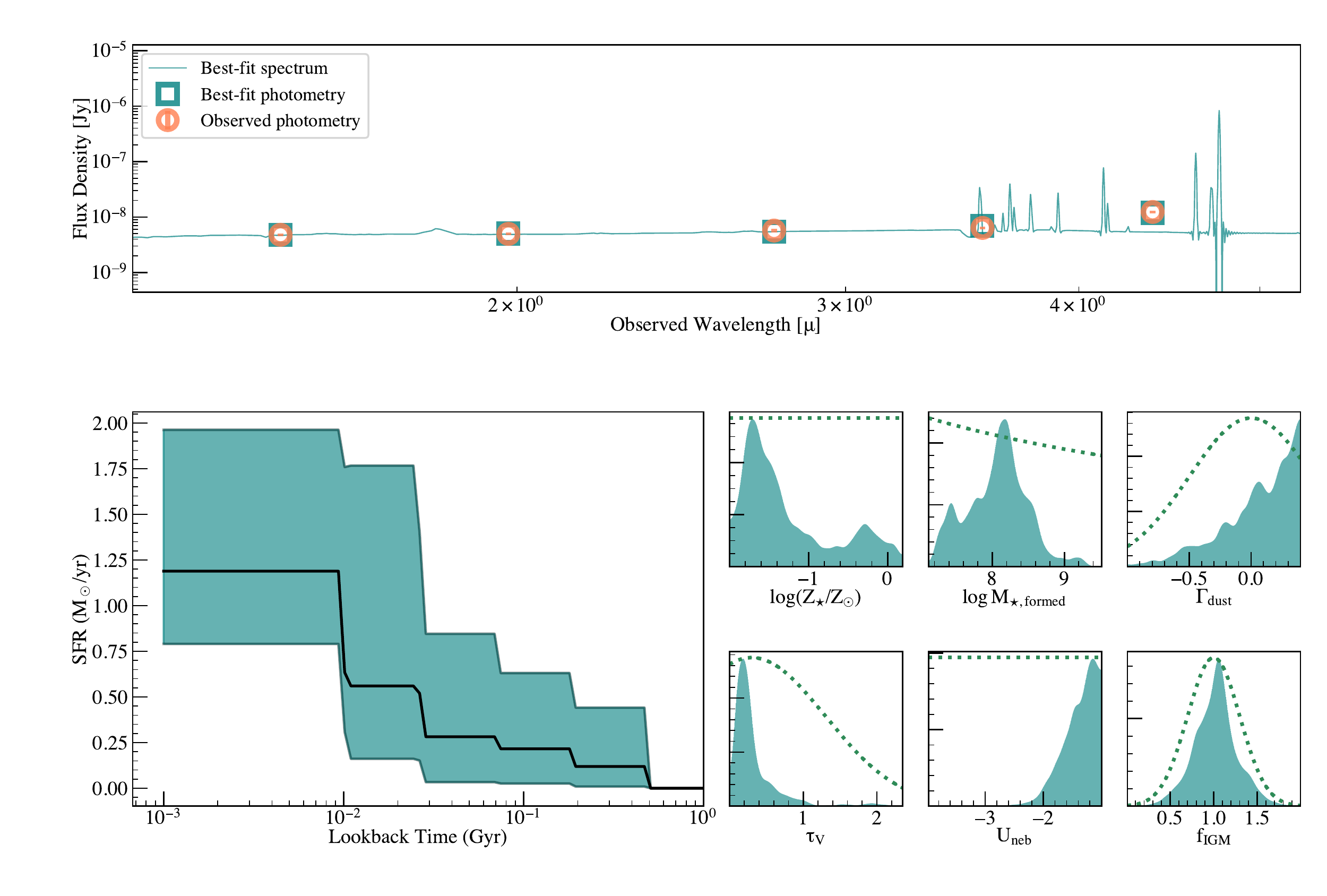}
    \caption{SED-fitting results for the SMACS0723--ID4590 galaxy.}
    \label{fig: 4590}
\end{figure*}

\begin{figure*}
    \centering
    \includegraphics[width=18cm]{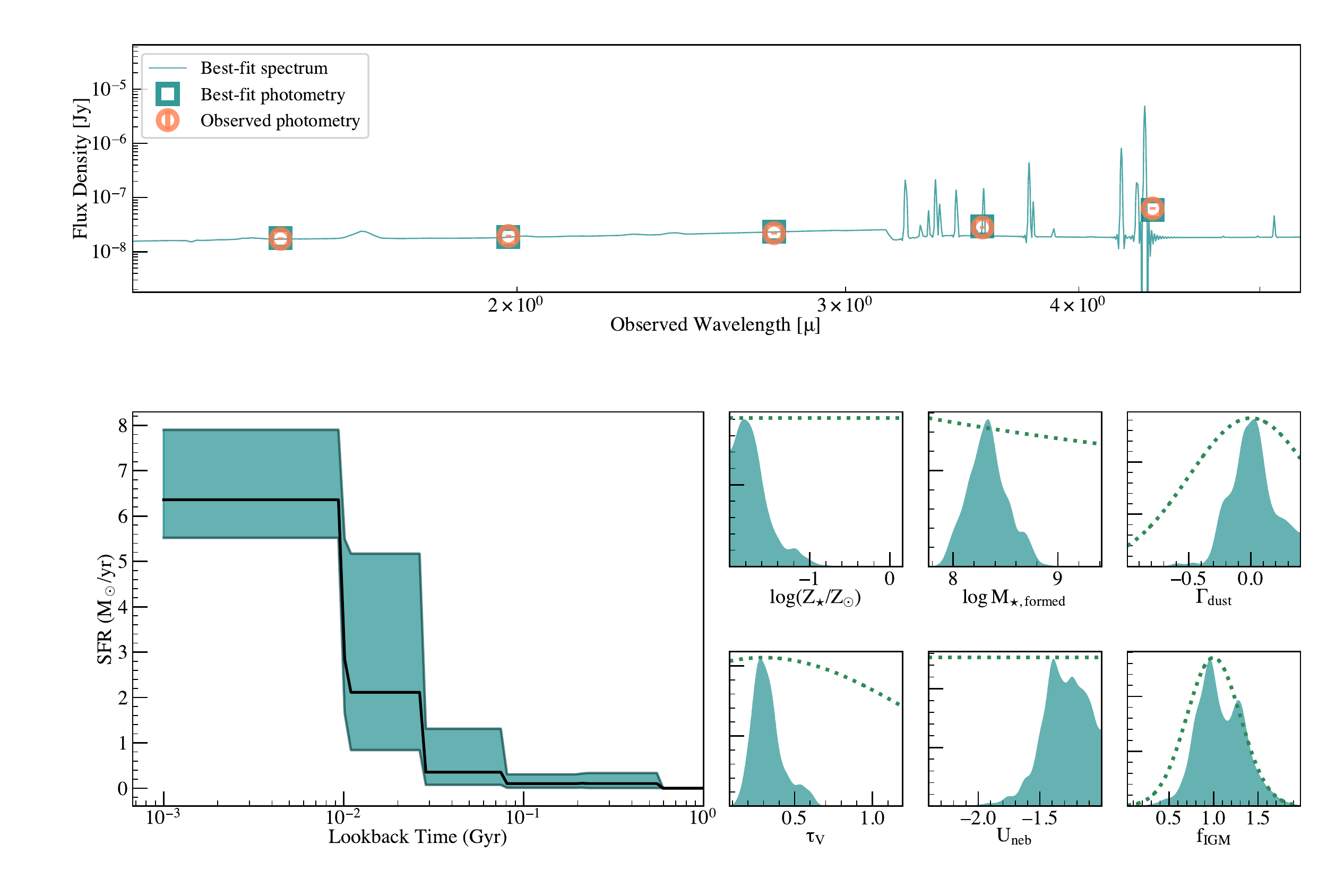}
    \caption{SED-fitting results for the SMACS0723--ID6355 galaxy.}
    \label{fig: 6355}
\end{figure*}

\begin{figure*}
    \centering
    \includegraphics[width=18cm]{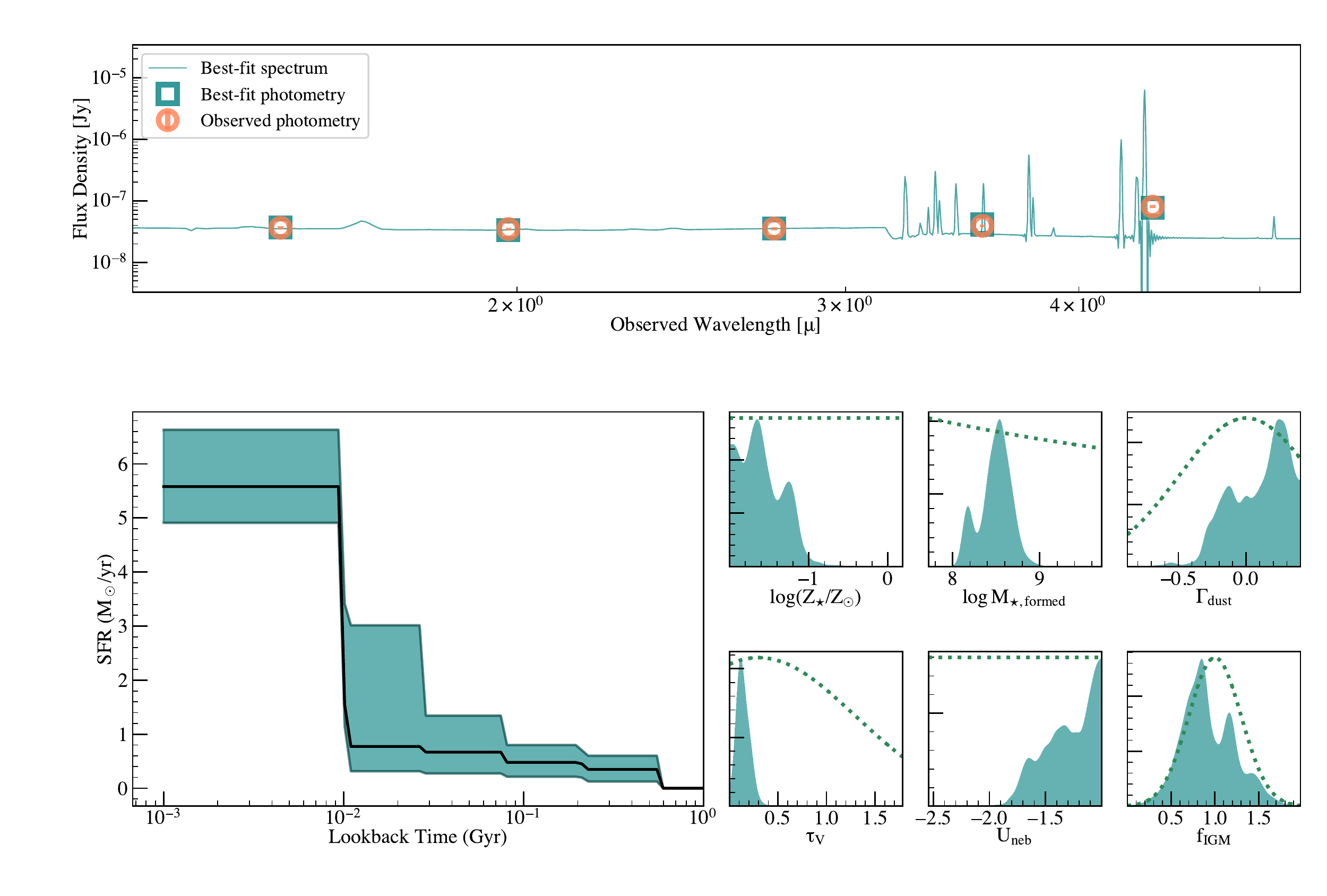}
    \caption{SED-fitting results for the SMACS0723--ID10612 galaxy.}
    \label{fig: 10612}
\end{figure*}

\begin{figure*}
    \centering
    \includegraphics[width=18cm]{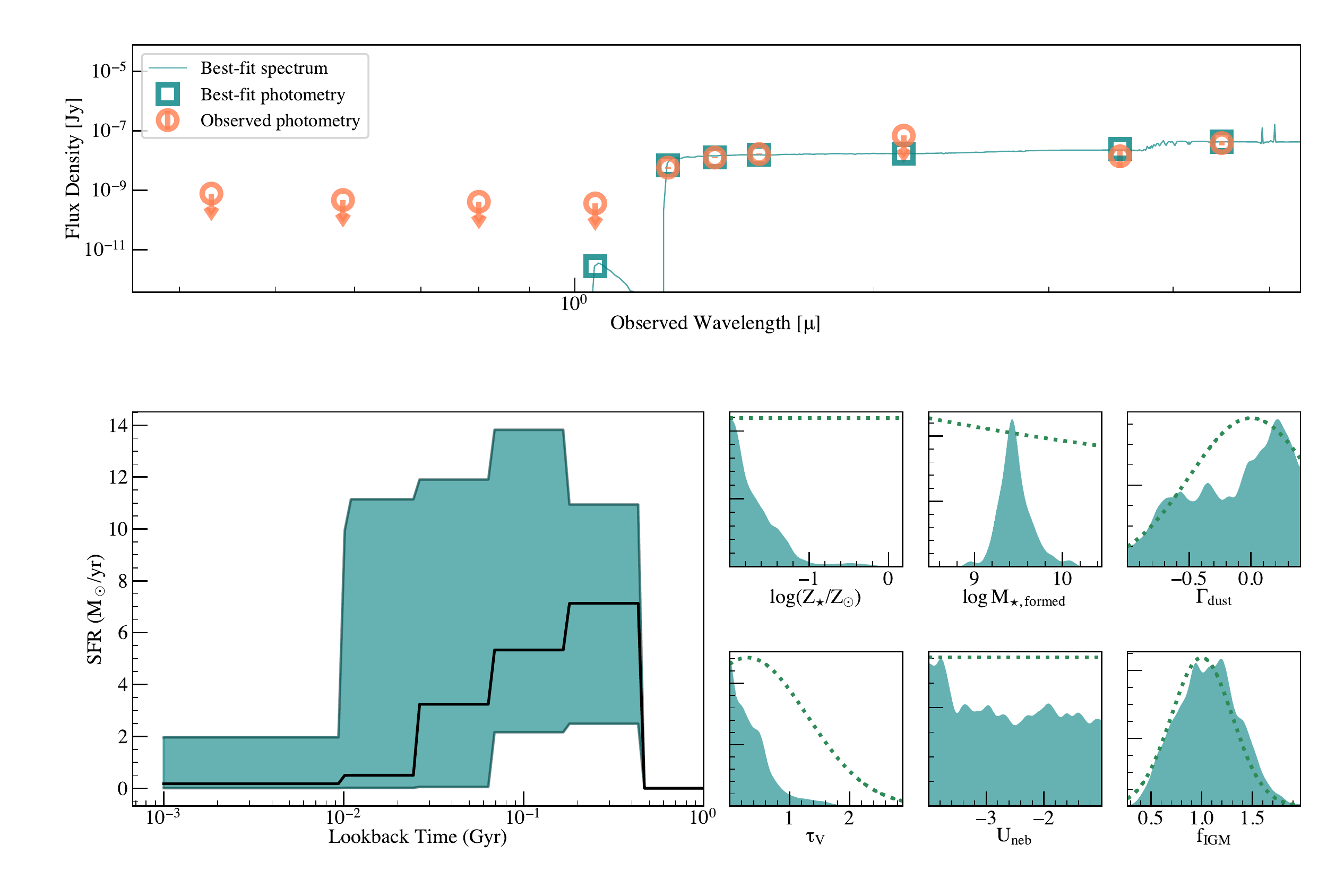}
    \caption{SED-fitting results for the MACS1149--JD1 galaxy.}
    \label{fig: MACS1149--JD1}
\end{figure*}

\begin{figure*}
    \centering
    \includegraphics[width=18cm]{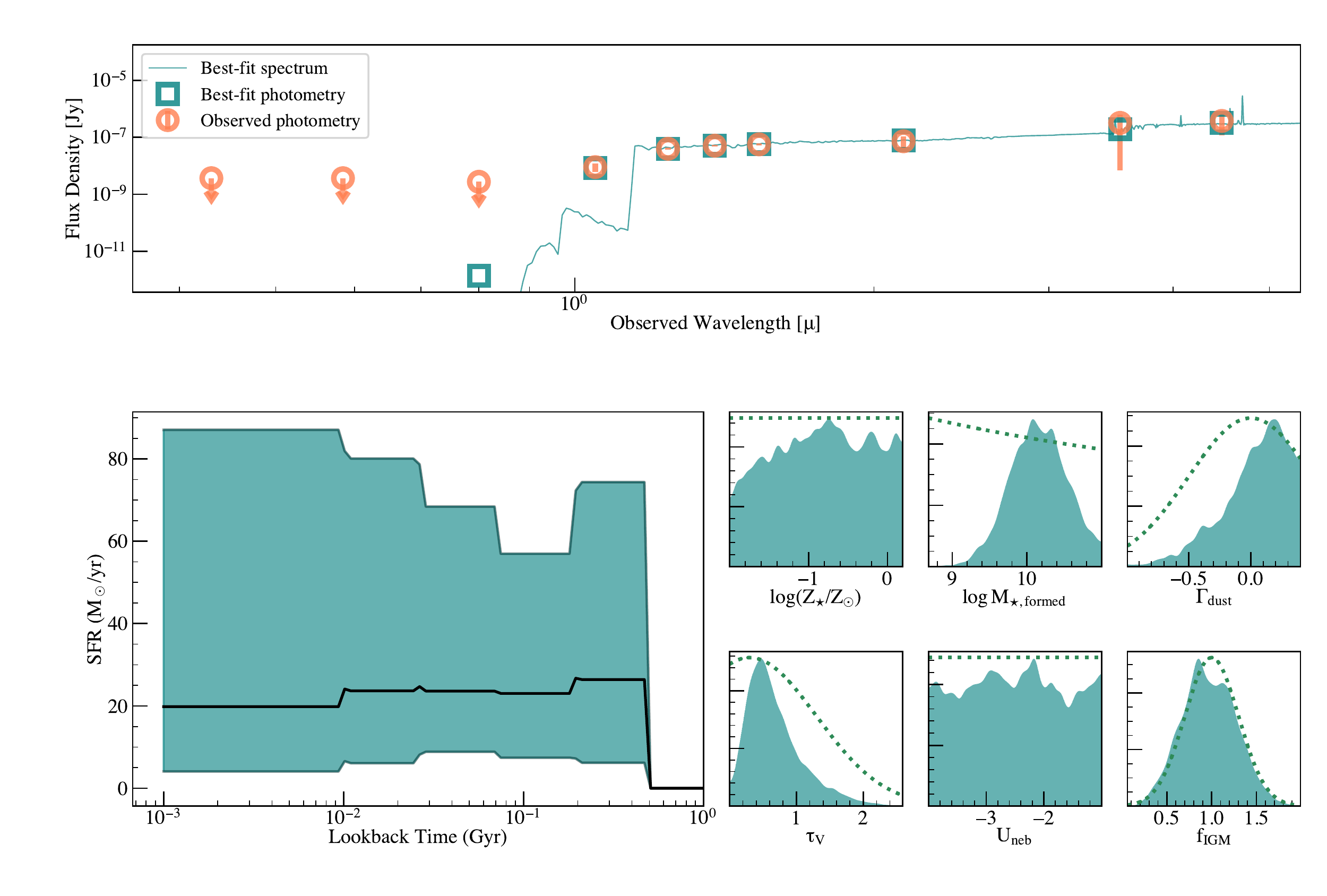}
    \caption{SED-fitting results for the A2744--YD4 galaxy.}
    \label{fig: A2744--YD4}
\end{figure*}

\begin{figure*}
    \centering
    \includegraphics[width=18cm]{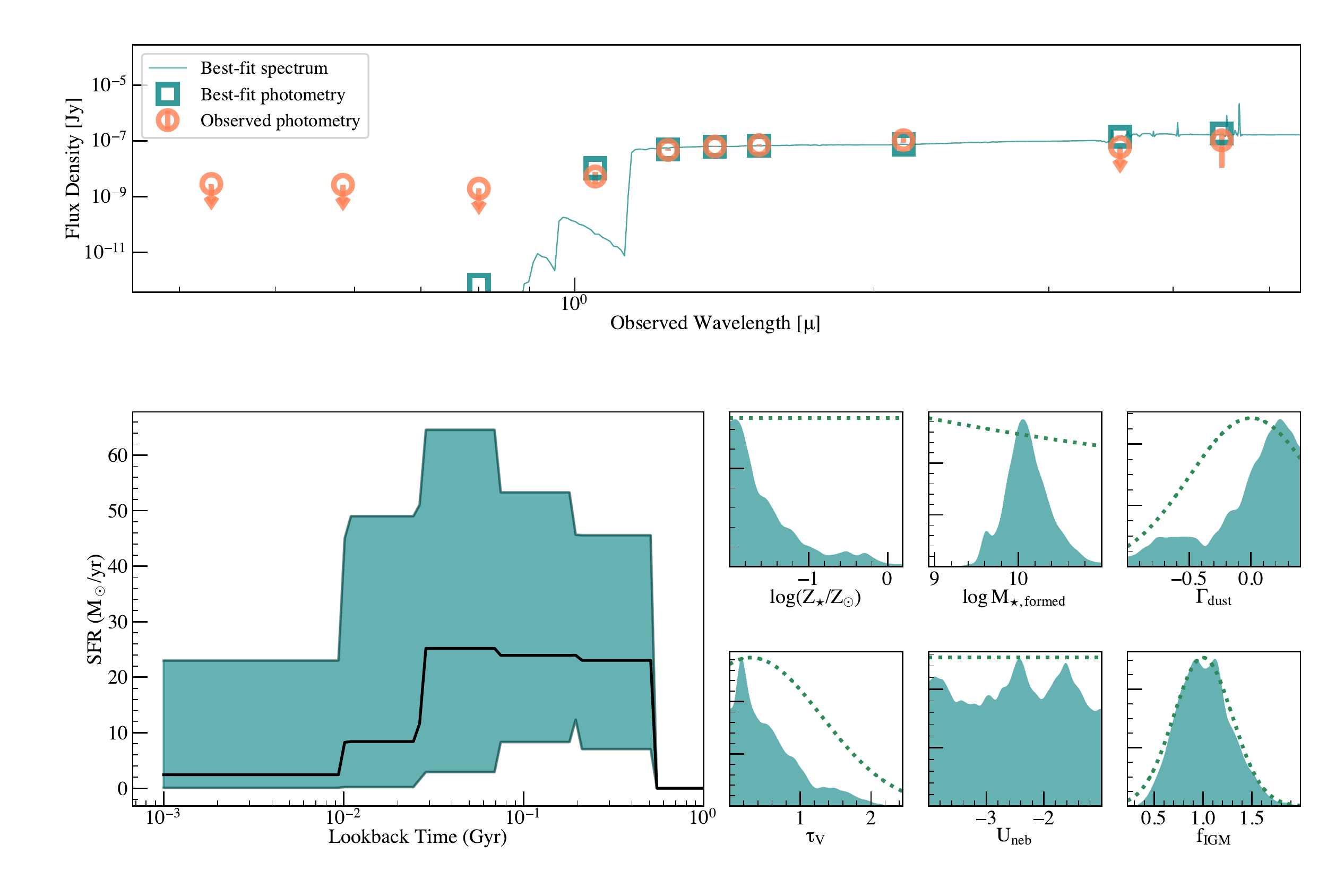}
    \caption{SED-fitting results for the MACS0416--Y1 galaxy.}
    \label{fig: MACS0416--Y1}
\end{figure*}

\begin{figure*}
    \centering
    \includegraphics[width=18cm]{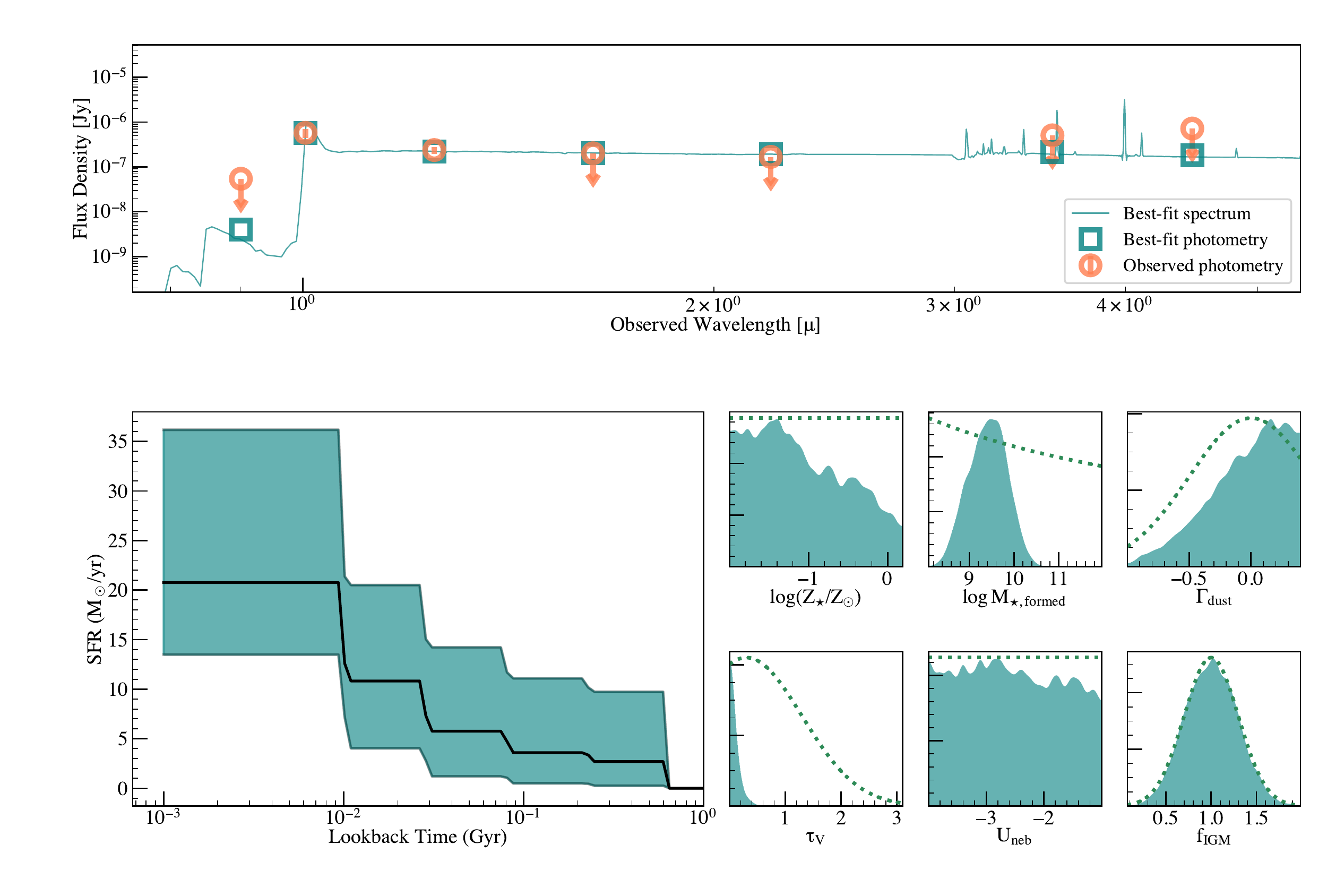}
    \caption{SED-fitting results for the SXDF--NB1006--2 galaxy.}
    \label{fig: SXDF--NB1006--2}
\end{figure*}

\begin{figure*}
    \centering
    \includegraphics[width=18cm]{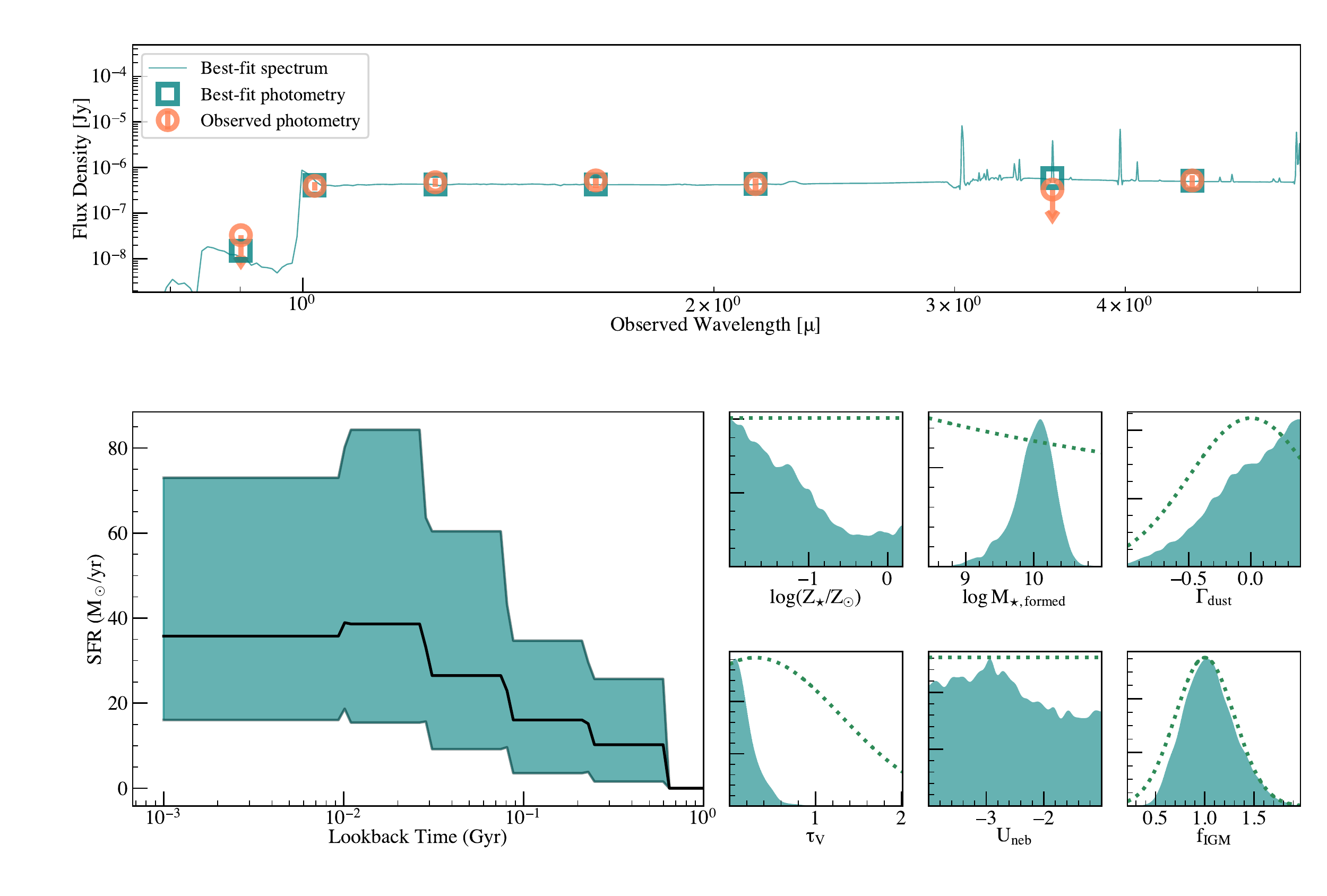}
    \caption{SED-fitting results for the B14--65666 galaxy.}
    \label{fig: B14--65666}
\end{figure*}

\end{document}